\documentclass[twocolumn,showpacs,showkeys,preprintnumbers,amsmath,amssymb,floatfix,aps]{revtex4}
\usepackage[dvips]{graphicx}
\usepackage{dcolumn}
\usepackage{rotating}
 
\begin{document}

\preprint{Phys. Rev. C}
 
\title {Separated Structure Functions for Exclusive $K^+\Lambda$ and
$K^+\Sigma^0$ Electroproduction at 5.5~GeV with CLAS}

\newcounter{univ_counter}
\setcounter{univ_counter} {0}

\addtocounter{univ_counter} {1} 
\edef\ANL{$^{\arabic{univ_counter}}$ }

\addtocounter{univ_counter} {1} 
\edef\ASU{$^{\arabic{univ_counter}}$ }

\addtocounter{univ_counter} {1} 
\edef\CSUDH{$^{\arabic{univ_counter}}$ }

\addtocounter{univ_counter} {1} 
\edef\CANISIUS{$^{\arabic{univ_counter}}$ }

\addtocounter{univ_counter} {1} 
\edef\CMU{$^{\arabic{univ_counter}}$ }

\addtocounter{univ_counter} {1} 
\edef\CUA{$^{\arabic{univ_counter}}$ }

\addtocounter{univ_counter} {1} 
\edef\SACLAY{$^{\arabic{univ_counter}}$ }

\addtocounter{univ_counter} {1} 
\edef\CNU{$^{\arabic{univ_counter}}$ }

\addtocounter{univ_counter} {1} 
\edef\UCONN{$^{\arabic{univ_counter}}$ }

\addtocounter{univ_counter} {1} 
\edef\EDINBURGH{$^{\arabic{univ_counter}}$ }

\addtocounter{univ_counter} {1} 
\edef\FU{$^{\arabic{univ_counter}}$ }

\addtocounter{univ_counter} {1} 
\edef\FIU{$^{\arabic{univ_counter}}$ }

\addtocounter{univ_counter} {1} 
\edef\FSU{$^{\arabic{univ_counter}}$ }

\addtocounter{univ_counter} {1} 
\edef\Genova{$^{\arabic{univ_counter}}$ }

\addtocounter{univ_counter} {1} 
\edef\GWUI{$^{\arabic{univ_counter}}$ }

\addtocounter{univ_counter} {1} 
\edef\ISU{$^{\arabic{univ_counter}}$ }

\addtocounter{univ_counter} {1} 
\edef\INFNFE{$^{\arabic{univ_counter}}$ }

\addtocounter{univ_counter} {1} 
\edef\INFNFR{$^{\arabic{univ_counter}}$ }

\addtocounter{univ_counter} {1} 
\edef\INFNGE{$^{\arabic{univ_counter}}$ }

\addtocounter{univ_counter} {1} 
\edef\INFNRO{$^{\arabic{univ_counter}}$ }

\addtocounter{univ_counter} {1} 
\edef\ORSAY{$^{\arabic{univ_counter}}$ }

\addtocounter{univ_counter} {1} 
\edef\ITEP{$^{\arabic{univ_counter}}$ }

\addtocounter{univ_counter} {1} 
\edef\JMU{$^{\arabic{univ_counter}}$ }

\addtocounter{univ_counter} {1} 
\edef\KNU{$^{\arabic{univ_counter}}$ }

\addtocounter{univ_counter} {1} 
\edef\LPSC{$^{\arabic{univ_counter}}$ }

\addtocounter{univ_counter} {1} 
\edef\UNH{$^{\arabic{univ_counter}}$ }

\addtocounter{univ_counter} {1} 
\edef\NSU{$^{\arabic{univ_counter}}$ }

\addtocounter{univ_counter} {1} 
\edef\OHIOU{$^{\arabic{univ_counter}}$ }

\addtocounter{univ_counter} {1} 
\edef\ODU{$^{\arabic{univ_counter}}$ }

\addtocounter{univ_counter} {1} 
\edef\RPI{$^{\arabic{univ_counter}}$ }

\addtocounter{univ_counter} {1} 
\edef\ROMAII{$^{\arabic{univ_counter}}$ }

\addtocounter{univ_counter} {1} 
\edef\MSU{$^{\arabic{univ_counter}}$ }

\addtocounter{univ_counter} {1} 
\edef\SCAROLINA{$^{\arabic{univ_counter}}$ }

\addtocounter{univ_counter} {1} 
\edef\JLAB{$^{\arabic{univ_counter}}$ }

\addtocounter{univ_counter} {1} 
\edef\UTFSM{$^{\arabic{univ_counter}}$ }

\addtocounter{univ_counter} {1} 
\edef\GLASGOW{$^{\arabic{univ_counter}}$ }

\addtocounter{univ_counter} {1} 
\edef\VIRGINIA{$^{\arabic{univ_counter}}$ }

\addtocounter{univ_counter} {1} 
\edef\WM{$^{\arabic{univ_counter}}$ }

\addtocounter{univ_counter} {1} 
\edef\YEREVAN{$^{\arabic{univ_counter}}$ }
 

\author{ 
D.S.~Carman,\JLAB\
K.~Park,\JLAB\
B.A.~Raue,\FIU\
K.P. ~Adhikari,\ODU\
D.~Adikaram,\ODU\
M.~Aghasyan,\INFNFR\
M.J.~Amaryan,\ODU\
M.D.~Anderson,\GLASGOW\
S. ~Anefalos~Pereira,\INFNFR\
M.~Anghinolfi,\INFNGE
% ...
\
\\
(CLAS Collaboration)
} 

\affiliation{\ANL Argonne National Laboratory, Argonne, Illinois 60439}
\affiliation{\ASU Arizona State University, Tempe, Arizona 85287}
\affiliation{\CSUDH California State University, Dominguez Hills, Carson, California 90747}
\affiliation{\CANISIUS Canisius College, Buffalo, New York 14208}
\affiliation{\CMU Carnegie Mellon University, Pittsburgh, Pennsylvania 15213}
\affiliation{\CUA Catholic University of America, Washington, D.C. 20064}
\affiliation{\SACLAY CEA, Centre de Saclay, Irfu/Service de Physique Nucl\'eaire, 91191 Gif-sur-Yvette, France}
\affiliation{\CNU Christopher Newport University, Newport News, Virginia 23606}
\affiliation{\UCONN University of Connecticut, Storrs, Connecticut 06269}
\affiliation{\EDINBURGH Edinburgh University, Edinburgh EH9 3JZ, United Kingdom}
\affiliation{\FU Fairfield University, Fairfield, Connecticut 06824}
\affiliation{\FIU Florida International University, Miami, Florida 33199}
\affiliation{\FSU Florida State University, Tallahassee, Florida 32306}
\affiliation{\Genova Universit$\grave{a}$ di Genova, 16146 Genova, Italy}
\affiliation{\GWUI The George Washington University, Washington, D.C. 20052}
\affiliation{\ISU Idaho State University, Pocatello, Idaho 83209}
\affiliation{\INFNFE INFN, Sezione di Ferrara, 44100 Ferrara, Italy}
\affiliation{\INFNFR INFN, Laboratori Nazionali di Frascati, 00044 Frascati, Italy}
\affiliation{\INFNGE INFN, Sezione di Genova, 16146 Genova, Italy}
\affiliation{\INFNRO INFN, Sezione di Roma Tor Vergata, 00133 Rome, Italy}
\affiliation{\ORSAY Institut de Physique Nucl\'eaire ORSAY, Orsay, France}
\affiliation{\ITEP Institute of Theoretical and Experimental Physics, Moscow, 117259, Russia}
\affiliation{\JMU James Madison University, Harrisonburg, Virginia 22807}
\affiliation{\KNU Kyungpook National University, Daegu 702-701, Republic of Korea}
\affiliation{\LPSC LPSC, Universite Joseph Fourier, CNRS/IN2P3, INPG, Grenoble, France}
\affiliation{\UNH University of New Hampshire, Durham, New Hampshire 03824}
\affiliation{\NSU Norfolk State University, Norfolk, Virginia 23504}
\affiliation{\OHIOU Ohio University, Athens, Ohio  45701}
\affiliation{\ODU Old Dominion University, Norfolk, Virginia 23529}
\affiliation{\RPI Rensselaer Polytechnic Institute, Troy, New York 12180}
\affiliation{\ROMAII Universita' di Roma Tor Vergata, 00133 Rome, Italy}
\affiliation{\MSU Skobeltsyn Nuclear Physics Institute, 119899 Moscow, Russia}
\affiliation{\SCAROLINA University of South Carolina, Columbia, South Carolina 29208}
\affiliation{\JLAB Thomas Jefferson National Accelerator Facility, Newport News, Virginia 23606}
\affiliation{\UTFSM Universidad T\'{e}cnica Federico Santa Mar\'{i}a, Casilla 110-V Valpara\'{i}so, Chile}
\affiliation{\GLASGOW University of Glasgow, Glasgow G12 8QQ, United Kingdom}
\affiliation{\VIRGINIA University of Virginia, Charlottesville, Virginia 22901}
\affiliation{\WM College of William and Mary, Williamsburg, Virginia 23187}
\affiliation{\YEREVAN Yerevan Physics Institute, 375036 Yerevan, Armenia}

\date{\today}

\begin{abstract}
We report measurements of the exclusive electroproduction of $K^+\Lambda$ and $K^+\Sigma^0$ final states 
from an unpolarized proton target using the CLAS detector at the Thomas Jefferson National Accelerator 
Facility. The separated structure functions $\sigma_U$, $\sigma_{LT}$, $\sigma_{TT}$, and $\sigma_{LT'}$ 
were extracted from the $\Phi$-dependent differential cross sections acquired with a longitudinally 
polarized 5.499~GeV electron beam. The data span a broad range of momentum transfers $Q^2$ from 1.4 to 
3.9~GeV$^2$, invariant energy $W$ from threshold to 2.6~GeV, and nearly the full center-of-mass angular 
range of the kaon. The separated structure functions provide an unprecedented data sample, which in 
conjunction with other meson photo- and electroproduction data, will help to constrain the higher-level
analyses being performed to search for missing baryon resonances.
\end{abstract}


\maketitle


\newpage

\section{Introduction}
\label{intro}

A complete mapping of the nucleon excitation spectrum is the key to a detailed understanding of the 
effective degrees of freedom of the nucleon and its associated dynamics. The most comprehensive predictions 
of this spectrum have come from various implementations of the constituent quark model incorporating broken 
SU(6) symmetry~\cite{burkert}. Additional dynamical contributions from gluonic excitations in the wavefunction 
may also play a central role~\cite{dudek}  and resonances may be dynamically generated through baryon-meson 
interactions~\cite{oset}. Quark model calculations of the nucleon spectrum have predicted more states than 
have been seen experimentally~\cite{capstick}. This has been termed the ``missing'' resonance problem, and 
the existence of these states is tied in directly with the underlying degrees of freedom of the nucleon that 
govern hadronic production at moderate energies~\cite{isgur}.

Ideally we should expect that the fundamental theory that governs the strong interaction, Quantum 
Chromodynamics (QCD), should provide a reliable prediction of the nucleon excitation spectrum. However, 
due to the non-perturbative nature of QCD at these energies, this expectation has not yet been fully 
realized. There has been notable recent progress in calculations of QCD on the lattice that has led to 
predictions of the nucleon excitation spectrum with dynamical quarks, albeit with unphysical pion masses 
\cite{buluva}. Calculations with improved actions, larger volumes, and smaller quark masses continue to 
progress.

In parallel, the development of coupled-channel models, such as those developed by the groups 
at Bonn-Gatchina~\cite{bonngat1,bonngat2}, Giessen~\cite{penner}, J\"{u}lich~\cite{julich}, and EBAC
\cite{ebac1}, have made significant progress toward deconvoluting the nucleon spectrum. These multi-channel 
partial wave analyses have employed partial wave fits from SAID~\cite{said} based on $\pi N$ elastic data 
to determine the properties of most $N^*$ and $\Delta^*$ resonances listed in the Particle Data Group 
(PDG)~\cite{pdg}. Further critical information on the decay modes was obtained by including the inelastic 
reactions $\pi N \to \eta N$, $K\Lambda$, $K\Sigma$, and $\pi \pi N$.

Recently the data landscape has undergone significant change with the publication of a vast amount 
of precision data in the photoproduction sector from JLab, SPring-8, MAMI, Bonn, and GRAAL. 
Data sets spanning a broad angular and energy range for $\gamma p \to p \pi^0$, $n \pi^+$, $p \eta$, 
$p \pi^0 \pi^0$, $p \pi^+ \pi^-$, $p \pi^0 \eta$, $K^+\Lambda$, and $K^+\Sigma^0$ have provided high
precision differential cross sections and polarization observables. Furthermore, new observables with 
polarized beams on both polarized proton and neutron targets have recently been acquired at several 
facilities and will be published over the next several years.

In the $K^+\Lambda$ and $K^+\Sigma^0$ electroproduction sector, dramatic changes to the world's database
occurred with the publications from the CLAS Collaboration. These include (i) beam-recoil transferred 
polarization for $K^+\Lambda$~\cite{carman_1} and for $K^+\Lambda$ and $K^+\Sigma^0$~\cite{carman_2}, (ii)
separated structure functions $\sigma_U = \sigma_T + \epsilon \sigma_L$, $\sigma_{LT}$, and $\sigma_{TT}$ 
for $K^+\Lambda$ and $K^+\Sigma^0$, as well as $\sigma_T$ and $\sigma_L$~\cite{5st}, and (iii) polarized 
structure function $\sigma_{LT'}$ for $K^+\Lambda$~\cite{sltp}.

This paper now adds to and extends this database with the largest data set ever acquired in these
kinematics for polarized electrons on an unpolarized proton target. This work includes measurements of 
the separated structure functions $\sigma_U$, $\sigma_{LT}$, $\sigma_{TT}$, and $\sigma_{LT'}$ for 
the $K^+\Lambda$ and $K^+\Sigma^0$ final states at a beam energy of 5.499~GeV, spanning $W$ from 
threshold to 2.6~GeV, $Q^2$ from 1.4 to 3.9~GeV$^2$, and nearly the full center-of-mass angular range 
of the kaon. The full set of differential cross sections $d\sigma/d\Omega_K^*$ included in this work 
consists of 480 (450) bins in $Q^2$, $W$, and $\cos \theta_K^*$ for the $K^+\Lambda$ ($K^+\Sigma^0$) 
final state and 3840 (3600) data points in $Q^2$, $W$, $\cos \theta_K^*$, and $\Phi$ for 
$K^+\Lambda$ ($K^+\Sigma^0$). 

The organization for this paper is as follows. In Section~\ref{theory}, the different 
theoretical models that are compared against the data are briefly described. In Section~\ref{formalism}, 
the relevant formalism for the expression of the electroproduction cross sections and separated structure 
functions is introduced. Section~\ref{analysis} details the experimental setup and describes all analysis 
cuts and corrections to the data. Section~\ref{systematics} details the sources of systematic uncertainty on 
the measured cross sections and separated structure functions, which are presented in Section~\ref{results}
along with a series of Legendre polynomial fits to the structure function data. Finally, we present a summary 
of this work and our conclusions in Section~\ref{conclusions}.

\section{Theoretical Models}
\label{theory}

To date the PDG lists only four $N^*$ states, $N(1650)1/2^-$, $N(1710)1/2^+$, $N(1720)3/2^+$, and $N(1900)3/2^+$,
with known couplings to $K\Lambda$ and no $N^*$ states are listed that couple to $K\Sigma$~\cite{pdg}; only a 
single $\Delta^*$ state, $\Delta(1920)3/2^+$, is listed with coupling strength to $K\Sigma$. The branching
ratios to $KY$ provided for these states are typically less than 10\% with uncertainties of the size of the measured
coupling. While the relevance of this core set of $N^*$ states in the $\gamma^{(*)} p \to K^+ \Lambda$ reaction 
has long been considered a well-established fact, this set of states falls short of reproducing the experimental 
results below $W$=2~GeV. Furthermore, recent analyses~\cite{sarantsev,shklyar} have called the importance of the 
$N(1710)1/2^+$ state into question. 

Beyond the core set of $N^*$ states, the PDG lists the $N(1900)3/2^+$ state as the sole established $N^*$ 
near 1900~MeV. However, with a 500-MeV width quoted by some measurements, it is unlikely that this state by 
itself could explain the $K^+\Lambda$ cross sections below $W$=2~GeV, unless its parameters are significantly 
different than those given by the PDG. Recent analyses~\cite{mart1,nikonov} have shown this state to be necessary 
to describe the CLAS beam-recoil polarization data~\cite{bradford}. Note that the $N(1900)3/2^+$ state is predicted 
by symmetric quark models and its existence is not expected in diquark models. In the recent fits of 
$\gamma p \to K^+ \Sigma^0$ data, all $N^*$ resonances found to be necessary to fit the $K^+\Lambda$ data have 
been included. However, the existing $K^+\Sigma^0$ database is smaller than the $K^+\Lambda$ database, with 
significantly larger statistical uncertainties. 

A recent development in understanding the $N^*$ spectrum was provided by the Bonn-Gatchina coupled-channel 
partial wave analysis of the hadronic $\pi N$ channels and the photoproduced $\gamma p$ channels~\cite{bonngat1}.
This work presents an up-to-date listing of pole parameters and branching fractions for all $N^*$ and $\Delta^*$ 
states up to $\sim$2~GeV with uncertainties at the level of a few percent. That analysis provided a list of
(i) six $N^*$ states with coupling to $K\Lambda$, $N(1650)1/2^-$, $N(1710)1/2^+$, $N(1875)3/2^-$, $N(1880)1/2^+$, 
$N(1895)1/2^-$, $N(1900)3/2^+$, (ii) five $N^*$ states with coupling to $K\Sigma$, $N(1875)3/2^-$, 
$N(1880)1/2^+$, $N(1895)1/2^-$, $N(1900)3/2^+$, $N(2060)5/2^-$, and (iii) four $\Delta^*$ states with coupling to
$K\Sigma$, $\Delta(1900)1/2^-$, $\Delta(1910)1/2^+$, $\Delta(1920)3/2^+$, $\Delta(1950)7/2^+$. For more on this
list of states that couple to $K\Lambda$ and $K\Sigma$, see Ref.~\cite{klempt}.

The findings of Ref.~\cite{bonngat1} are based on a significant amount of precision experimental data and the 
sophisticated coupled-channel fitting algorithms. However, in general, the issue of how to extract 
nucleon resonance content from open strangeness reactions is a long-standing question. Various analyses 
have led to very different conclusions concerning the set of resonances that contribute~(e.g. compare 
results from Refs.~\cite{nikonov}, \cite{diaz}, and \cite{martsul}, as well as the statements made regarding
the resonant set from Ref.~\cite{bonngat1}). Furthermore, lack of sufficient experimental information, incomplete 
kinematic coverage, and underestimated systematics are still responsible for inconsistencies among the different 
models that fit the data to extract the contributing resonances and their properties~\cite{bonngat2}.

The indeterminacy for the open strangeness channels is in contrast to the pionic channels, where the contributing 
resonances can be more reliably identified by means of a partial wave analysis for $W < 2$~GeV. In open strangeness 
channels, this technique is less powerful as the non-resonant background contributions are a much larger fraction of 
the overall response. Several groups have stressed that the importance of the background contributions calls for a 
framework that accounts for both the resonant and non-resonant processes and that provides for a means to constrain 
both of these classes of reaction mechanisms independently~\cite{corthals,maxwell1}.

While there have been a number of publications of precision cross sections and spin observables for both 
the photo- and electroproduction reactions, the vast majority of the theoretical effort has focused on 
fitting just the photoproduction data. Although $KY$ photoproduction is easier to treat theoretically 
than $KY$ electroproduction, and is thus more amenable to a detailed quantitative analysis, the 
electroproduction reaction is potentially a much richer source of information concerning hadronic and 
electromagnetic interactions. The electroproduction observables have been shown to yield important 
complementary insights~\cite{corthals}. Some of the most important aspects of electroproduction include:

\begin{itemize}
\item The data are sensitive to the internal structure of baryon resonances through the $Q^2$ dependence of the 
electromagnetic form factors of the intermediate hadronic resonances associated with the strangeness production 
mechanism~\cite{bonngat2}.

\item The structure functions are particularly powerful to gain control over the parameterization of the 
background diagrams~\cite{janssen}.

\item Studies of finite $Q^2$ processes are sensitive to both transverse and longitudinal virtual
photon couplings, in contrast to the purely transverse response probed in the photoproduction
reactions.

\item The longitudinal/transverse interference structure functions provide signatures of interfering partial 
wave strengths that are often dramatic and have been shown to be useful for differentiating between models 
of the production amplitudes~\cite{5st,sltp,mart2}.

\item The beam-recoil transferred polarizations in the $K^+\Lambda$ and $K^+\Sigma^0$ reactions, 
as well as the recoil polarization in the $K^+\Lambda$ reaction, have been shown to provide important
new constraints to models that describe well the photoproduction data~\cite{carman_1,carman_2,indpol}.
\end{itemize}

At the medium energies of this work, perturbative QCD is not yet capable of providing predictions of 
differential cross sections. To understand the underlying physics, effective models must be employed that 
represent approximations to QCD. Ultimately, it will be most appropriate to compare the electroproduction 
measurements against the results of a full coupled-channel partial wave analysis that is constrained by 
fits to the available data. Although output from such models is expected in the electroproduction sector 
in the future~\cite{harry-lee,thoma}, as of now, these data have not yet been included in the fits. Thus 
comparisons of the electroproduction observables to single-channel models currently represent the best 
option to gain insight into the electroproduction realm.

This analysis highlights three different theoretical model approaches. The first is a traditional 
hadrodynamic model and the second is based on $K$ and $K^*$ Regge trajectory exchange. The third model,
a hybrid Regge plus resonance approach, amounts to a cross between the first two model types. Comparison of 
the different model predictions to the data can be used to provide indirect support for the existence of the 
different baryonic resonances and their branching ratios into the strange channels, as well as to improve 
constraints on the phenomenology of the different strangeness production reactions. The following subsections 
provide a brief description of the models included in this work.

\subsection{Hadrodynamic Model}

\begin{figure}[htbp]
\vspace{2.7cm}
\includegraphics{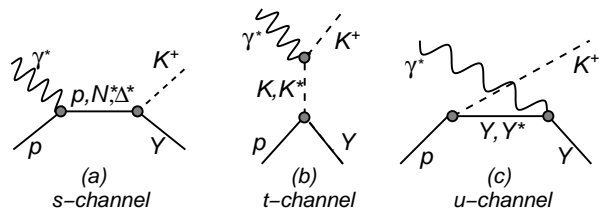}
\caption{Tree-level diagrams contributing to the $KY$ reactions: (a) $s$-channel exchanges, (b) $t$-channel 
exchanges, and (c) $u$-channel exchanges.}
\label{born}
\end{figure}

Hadrodynamic models provide a description of the reaction based on contributions from tree-level Born and 
extended Born terms in the $s$, $t$, and $u$ reaction channels (see Fig.~\ref{born}). The Born diagrams 
include the exchange of the proton, kaon, and ground-state hyperons, while the extended Born diagrams include 
the exchange of the associated excited states. This description of the interaction, which involves only 
first-order terms, is sensible as the incident and outgoing electrons interact rather weakly with the hadrons. 
A complete description of the physics processes requires taking into account all possible channels that could 
couple to the initial and final states, but the advantages of the tree-level approach are to limit complexity 
and to identify the dominant trends. The drawback in this class of models is that very different conclusions 
about the strengths of the contributing diagrams may be reached depending on which set of resonances a given 
model includes.

Maxwell {\it et al.}~\cite{maxwell1,maxwell2,maxwell3} have developed a tree-level effective Lagrangian model 
(referred to as MX) for $\gamma^{(*)} p \to K^+\Lambda$ that incorporates the well-established $s$-channel 
resonances up to 2.2~GeV with spins up to 5/2. The model also includes four $\Lambda$ $u$-channel states, 
$\Lambda(1405)1/2^-$, $\Lambda(1670)1/2^-$, $\Lambda(1820)5/2^+$, $\Lambda(1890)3/2^+$, four $\Sigma$ $u$-channel 
states, $\Sigma(1385)3/2^+$, $\Sigma(1775)5/2^-$, $\Sigma(1915)5/2^+$, $\Sigma(1940)3/2^-$, and the $K^*(892)$ and 
$K_1(1270)$ $t$-channel resonances.

The model was initially developed and fit to the available $\gamma p$ photoproduction data up to
$W$=2.3~GeV~\cite{maxwell3}. The most recent published version of the model~\cite{maxwell1} included 
fits to the available $K^+\Lambda$ separated structure function data from CLAS~\cite{5st}. An extension
of this model that also includes fits to the available CLAS $K^+\Lambda$ $\sigma_{LT'}$ data has been 
made available for this work as well. Overall the fits yield reasonable representations of both the photo- 
and electroproduction data. However, when compared to the results of a fit to the photoproduction data alone, 
the combined $\gamma p$ and $\gamma^* p$ fit yields significantly different coupling parameters for an 
equally good overall fit to the data. This indicates that the photoproduction data alone are not adequate to 
uniquely constrain effective Lagrangian models of electromagnetic strangeness production.

\subsection{Regge Model}

Our $KY$ electroproduction data are also compared to the Regge model from Guidal, Laget, and Vanderhaeghen
\cite{glv} (referred to as GLV). This calculation includes no baryon resonance terms at all. Instead, it is 
based only on gauge-invariant $t$-channel $K$ and $K^*$ Regge-trajectory exchange. It therefore provides a 
complementary basis for studying the underlying dynamics of strangeness production. It is important to note 
that the Regge approach has far fewer parameters compared to the hadrodynamic models. These include the $K$ 
and $K^*$ form factors and the coupling constants $g_{KYN}$ and $g_{K^*YN}$.

The GLV model was fit to higher-energy photoproduction data where there is little doubt of the dominance 
of kaon exchanges, and extrapolated down to JLab energies. An important feature of this model is the way 
gauge invariance is achieved for the kaonic $t$-channel exchanges by Reggeizing the $s$-channel nucleon 
pole contribution in the same manner as the $t$-channel diagrams. No counter terms need to be introduced 
to restore gauge invariance as is done in the hadrodynamic approach.

The GLV Regge model reasonably accounts for the strength in the CLAS $K^+\Lambda$ differential cross sections 
and separated structure functions~\cite{5st}. Although the reasonable performance of a pure Regge 
description in this channel suggests a $t$-channel dominated process, there are obvious discrepancies between 
the Regge predictions and the data, indicative of $s$-channel strength. In the $K^+\Sigma^0$ channel, the
same Regge description significantly underpredicts the differential cross sections and separated structure 
functions~\cite{5st}. The fact that the Regge model fares poorly when compared to the $K^+\Sigma^0$ 
data is indicative that this process has a much larger $s$-channel content compared to $K^+\Lambda$ production.

\subsection{Regge Plus Resonance Model}
\label{rpr-theory}

The final model included in this work was developed by the Ghent group~\cite{corthals}, and is based 
on a tree-level effective field model for $K^+\Lambda$ and $K^+\Sigma^0$ photoproduction from the proton. It 
differs from traditional isobar approaches in its description of the non-resonant diagrams, which involve 
the exchange of $K$ and $K^*$ Regge trajectories. A selection of $s$-channel resonances is then added to 
this background. This ``Regge plus resonance'' model (referred to as RPR) has the advantage that the 
background diagrams contain only a few parameters that are tightly constrained by high-energy data. 
Furthermore, the use of Regge propagators eliminates the need to introduce strong form factors in the 
background terms, thus avoiding the gauge-invariance issues associated with traditional effective Lagrangian 
models. 

In addition to the kaonic trajectories to model the $t$-channel background, the RPR model includes the 
same $s$-channel resonances as for the MX model below 2~GeV. The model does include several missing 
$N^*$ states at 1.9~GeV, $N(1900)3/2^-$, $N(1900)3/2^-$, and $N(1900)1/2^+$. The separated structure 
functions~\cite{5st,sltp} and beam-recoil transferred polarization data from CLAS~\cite{carman_2} were compared 
to model variants with either a $N(1900)3/2^-$ or a $N(1900)1/2^+$ state at 1.9~GeV. Only the $N(1900)3/2^-$ state 
assumption could be reconciled with the data, whereas the $N(1900)1/2^+$ option could clearly be rejected. In the 
$K^+\Sigma^0$ channel, four $\Delta^*$ states, $\Delta(1700)3/2^-$, $\Delta(1900)1/2^-$, $\Delta(1910)1/2^+$, and 
$\Delta(1920)3/2^+$, have been included.

In a new version of the RPR model (referred to as RPR-2011)~\cite{decruz}, several changes relative to the 
previous model version (referred to as RPR-2007)~\cite{corthals} are noteworthy. The main difference is the 
implementation of an unbiased model selection methodology based on Bayesian inference. This inference is used 
as a quantitative measure of whether the inclusion of a given set of $N^*$ states is justified by the data. 
Additionally, in this version of the model, the exchange of spin-3/2 resonances is described within a 
consistent interaction theory and the model has been extended to include the exchange of spin 5/2 resonances.

The Regge background amplitude of RPR-2007 is constrained by spectra above the resonance region ($W > 3$~GeV)
at forward angles ($\cos \theta_K^* > 0.35$). By extrapolating the resulting amplitude to smaller $W$, one
gets a parameter free background for the resonance region. The $s$-channel resonances are coherently added
to the background amplitude. RPR-2007 describes the data for forward-angle photo- and electroproduction
of $K^+\Lambda$ and $K^+\Sigma^0$. The resonance parameters of the RPR-2007 model are constrained to the 
$\cos \theta_K^* > 0.35$ data. The RPR-2011 model with the highest evidence has nine well-established $N^*$ 
states and the ``missing'' states at 1.9~GeV with quantum numbers $N(1900)3/2^-$ and $N(1900)1/2^+$, and has 
been fit to photoproduction data over the full $K^+$ center-of-mass (c.m.) angular range. Neither version of 
the model has been constrained by fits to any of the electroproduction data.

\section{Formalism}
\label{formalism}

\begin{figure}[htbp]
\vspace{5.6cm}
\includegraphics{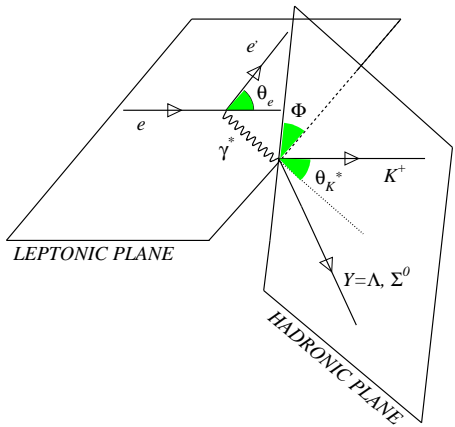}
\caption{Kinematics for $K^+Y$ electroproduction defining the angles $\theta_K^*$ and $\Phi$.} 
\label{fig-kin}
\end{figure}

In kaon electroproduction a beam of electrons with four-momentum $p_e = (E_e,\vec{p}_e\,)$ is 
incident upon a fixed proton target of mass $M_p$, and the outgoing scattered electron with momentum 
$p_{e'}=(E_{e'},\vec{p}_{e'}\,)$ and kaon with momentum $p_K=(E_K,\vec{p}_K)$ are measured. The 
cross section for the exclusive $K^+Y$ final state is then differential in the scattered electron 
momentum and kaon direction. Under the assumption of single-photon exchange, where the virtual photon 
has four-momentum $q=p_e-p_{e'}=(\nu,\vec{q}\,)$, this can be expressed as the product of an 
equivalent flux of virtual photons and the $\gamma^* p$ c.m. virtual photoabsorption cross section as:
\begin{equation}
\frac{d^5\sigma}{dE_{e'} d\Omega_{e'} d\Omega_K^*} = \Gamma \frac{d^2\sigma_v}{d\Omega_K^*},
\end{equation}
\noindent
where the virtual photon flux factor $\Gamma$ depends upon only the electron scattering process. 
After integrating over the azimuthal angle of the scattered electron, the absorption cross section 
can be expressed in terms of the variables $Q^2$, $W$, $\theta_K^*$, and $\Phi$, where $q^2=-Q^2$ is 
the squared four-momentum of the virtual photon, $W = \sqrt{M_p^2+2M_p\nu-Q^2}$ is the total hadronic 
energy in the c.m. frame, $\theta_K^*$ is the c.m. kaon angle relative to the virtual photon direction, 
and $\Phi$ is the angle between the leptonic and hadronic production planes. A schematic illustration 
of electron scattering off a proton target, producing a final state electron, $K^+$, and hyperon $Y$ 
is shown in Fig.~\ref{fig-kin}. 

Introducing the appropriate Jacobian, the form of the cross section can be rewritten as:
\begin{equation}
\frac{d^4\sigma}{dQ^2 dW d\Omega_K^*} = \Gamma_v  \frac{d^2\sigma_v}{d\Omega_K^*},
\end{equation}
\noindent
where
\begin{equation}
\label{eq:flux}
\Gamma_v = \frac{\alpha}{4\pi}\frac{W}{M_p^2 E^2}\frac{W^2-M_p^2}{Q^2} \frac{1}{1-\epsilon}
\end{equation}
\noindent
is the flux of virtual photons (using the definition from Ref.~\cite{akerlof}), 
\begin{equation}
\epsilon=\left(1+2\frac{\nu^2}{Q^2}\tan^2{\frac{\theta_{e'}}{2}} \right)^{-1}
\end{equation}
\noindent
is the polarization parameter of the virtual photon, and $\theta_{e'}$ is the electron scattering 
angle in the laboratory frame.

For the case of an unpolarized electron beam (helicity $h$=0) with no target or recoil polarizations, the 
virtual photon cross section can be written (using simplifying notation for the differential cross section)
as:
\begin{multline}
\label{sigma0}
\frac{d\sigma}{d\Omega_K^*} (h=0) \equiv \sigma_0 = \sigma_U + \epsilon \sigma_{TT} \cos 2\Phi \\
+ \sqrt{\epsilon(1 + \epsilon)} \sigma_{LT} \cos\Phi, 
\end{multline} 
\noindent
where $\sigma_i$ are the structure functions that measure the response of the hadronic system and $i=T$, 
$L$, $LT$, and $TT$ represents the transverse, longitudinal, and interference structure functions.
The structure functions are, in general, functions of $Q^2$, $W$, and $\theta_K^*$ only. In this work
the unseparated structure function is defined as $\sigma_U = \sigma_T + \epsilon \sigma_L$.

In contrast to the case of real photons, where there is only the purely transverse response, virtual photons 
allow longitudinal, transverse-transverse, and longitudinal-transverse interference terms to occur. Each of 
the structure functions is related to the coupling of the hadronic current to different combinations 
of the transverse and longitudinal polarization of the virtual photon. $\sigma_T$ is the differential cross 
section contribution for unpolarized transverse virtual photons. In the limit $Q^2 \to 0$, this term must 
approach the cross section for unpolarized real photons. $\sigma_L$ is the differential cross section contribution 
for longitudinally polarized virtual photons. $\sigma_{TT}$ and $\sigma_{LT}$ represent contributions to the cross 
section due to the interference of transversely polarized virtual photons and from transversely and longitudinally 
polarized virtual photons, respectively. 

For the case of a polarized electron beam with helicity $h$, the cross section form of Eq.(\ref{sigma0})
is modified to include an additional term:
\begin{equation}
\label{sig-hel}
\frac{d\sigma}{d\Omega_K^*} = \sigma_0 + h \sqrt{\epsilon(1 - \epsilon)} \sigma_{LT'} \sin\Phi. 
\end{equation} 
\noindent
The electron beam polarization produces a fifth structure function $\sigma_{LT'}$ that is related to the beam 
helicity asymmetry via:
\begin{equation}
\label{eq:sigltp}
A_{LT'} = \frac{\frac{d\sigma^+}{d\Omega_K^*} - \frac{d\sigma^-}{d\Omega_K^*}}
{\frac{d\sigma^+}{d\Omega_K^*} + \frac{d\sigma^-}{d\Omega_K^*}} = 
\frac{\sqrt{\epsilon(1-\epsilon)}\sigma_{LT'} \sin\Phi}{\sigma_0},
\end{equation}
\noindent
where the $\pm$ superscripts on $\frac{d\sigma}{d\Omega_K^*}$ correspond to the electron helicity states of $h=\pm 1$. 

The polarized structure function $\sigma_{LT'}$ is intrinsically different from the structure functions of the 
unpolarized cross section. This term is generated by the imaginary part of terms involving the interference 
between longitudinal and transverse components of the hadronic and leptonic currents, in contrast to 
$\sigma_{LT}$, which is generated by the real part of the same interference. $\sigma_{LT'}$ is non-vanishing 
only if the hadronic tensor is antisymmetric, which will occur in the presence of rescattering effects, 
interferences between multiple resonances, interferences between resonant and non-resonant processes, or even 
between non-resonant processes alone~\cite{boffi}. $\sigma_{LT'}$ could be non-zero even when $\sigma_{LT}$ is 
zero. When the reaction proceeds through a channel in which a single amplitude dominates, the longitudinal-transverse 
response will be real and $\sigma_{LT'}$ will vanish. Both $\sigma_{LT}$ and $\sigma_{LT'}$ are necessary to fully 
unravel the longitudinal-transverse response of the $K^+Y$ electroproduction reactions.

\section{Experiment Description and Data Analysis}
\label{analysis}

The measurement was carried out with the CEBAF Large Acceptance Spectrometer (CLAS)~\cite{mecking} located in 
Hall~B at JLab. The main magnetic field of CLAS is provided by six superconducting coils, which produce an 
approximately toroidal field in the azimuthal direction around the beam axis. The gaps between the cryostats 
are instrumented with six identical detector packages. Each sector consists of drift chambers (DC)~\cite{dcnim} 
for charged particle tracking, Cherenkov counters (CC)~\cite{ccnim} for electron identification, scintillator 
counters (SC)~\cite{scnim} for charged particle identification, and electromagnetic calorimeters (EC)~\cite{ecnim} 
for electron identification and detection of neutral particles. A 5-cm-long liquid-hydrogen target was located 
25~cm upstream of the nominal center of CLAS. The main torus was operated at 60\% of its maximum field value and 
had its polarity set such that negatively charged particles were bent toward the electron beam line. A totally 
absorbing Faraday cup located at the end of the beam line was used to determine the integrated beam charge passing 
through the target. 

The efficiency of detection and reconstruction for stable charged particles in the fiducial regions of CLAS is 
greater than 95\%. The solid angle coverage of CLAS is approximately 3$\pi$~sr. The polar angle coverage for 
electrons ranges from 8$^{\circ}$ to 45$^{\circ}$, while for hadrons it is from 8$^{\circ}$ to 140$^{\circ}$, 
with an angular resolution of $\delta \theta, \delta \phi$ of better than 2~mr. The CLAS detector was designed 
to track particles having momenta greater than roughly 200~MeV with a resolution $\delta p/p$ of about 1\%.

The data in this paper were collected as part of the CLAS e1f running period in 2003. The incident electron 
beam energy was 5.499~GeV. The live-time corrected integrated luminosity of this data set is 10.6~fb$^{-1}$. 
The data set contains $3.64 \times 10^5$ $e'K^+\Lambda$ events and $1.56 \times 10^5$ $e'K^+\Sigma^0$ 
events in the analysis bins included in this work. 

The data were taken at an average electron beam current of 7~nA at a luminosity of about
$10^{34}$~cm$^{-2}$s$^{-1}$. The event readout was triggered by a coincidence between a 
CC hit and an EC hit in a single sector, generating an event rate of $\sim$2~kHz. The electron beam was 
longitudinally polarized with polarization determined by a coincidence M{\o}ller polarimeter. The average
beam polarization was about 75\%.

This analysis sought to measure the differential cross sections for the electroproduction reactions
$ep \to e'K^+\Lambda$ and $ep \to e'K^+\Sigma^0$ in bins of $Q^2$, $W$, $\cos \theta_K^*$, 
and $\Phi$. Exploiting the $\Phi$ dependence of the differential cross sections $\sigma_0$ as given by
Eq.(\ref{sigma0}), a $\Phi$ fit in each bin of $Q^2$, $W$, and $\cos \theta_K^*$ provides the
separated structure functions $\sigma_U$, $\sigma_{LT}$, and $\sigma_{TT}$. Finally, a $\Phi$ fit
to the beam spin asymmetry as given by Eq.(\ref{eq:sigltp}) in each bin of $Q^2$, $W$, and $\cos \theta_K^*$ 
gives access to the polarized structure function $\sigma_{LT'}$.

\subsection{Differential Cross Section Determination}
\label{diff-cs}

The bin-centered differential cross section for each hyperon final state in each kinematic bin $i$ was 
computed using the form:
\begin{multline}
\label{dcs}
\frac{d\sigma_i}{d\Omega^*_K} = \frac{1}{\Gamma_v} \cdot \frac{1}{\left( \Delta Q^2 \Delta W
\Delta \cos \theta_K^* \Delta \Phi \right)} \cdot \\ \frac{R_i \cdot N_i \cdot BC_i}{\eta_i \cdot N_0} \cdot
\frac{1}{\left(N_A \rho t / A_w \right)}, 
\end{multline}
\noindent
where $\Gamma_v$ is the virtual photon flux factor computed according to Eq.(\ref{eq:flux}) for each
bin at the bin-averaged mean of the bin and $\Delta Q^2 \Delta W \Delta \cos \theta_K^* \Delta \Phi $ is the 
volume of each analysis bin computed using the bin sizes listed in Section~\ref{anal-pid} (the bin sizes 
are corrected for kinematic limits in the threshold $W$ bins). $R_i$ is the radiative
correction factor, $N_i$ is the background-subtracted $K^+\Lambda$ and $K^+\Sigma^0$ yield in each bin,
$BC_i$ is the factor that evolves the measured bin-averaged differential cross section over each bin
to a specific kinematic point within the $Q^2$, $W$, $\cos \theta_K^*$, $\Phi$ bin, and $\eta_i$ accounts for 
the detector geometrical acceptance and efficiency corrections. $N_0$ is the live-time corrected incident 
electron flux summed over all data runs included in this analysis determined from the Faraday Cup charge. For 
this experiment, the data acquisition live time ranged between 80 and 85\%. The incident electron flux was 
measured to be $N_0=9.807\times10^{16}$. Finally, $N_A \rho t / A_w$ represents the target number density, where
$N_A$ is Avogadro's number, $\rho$=0.07151~g/cm$^3$ is the target density, $t$=5.0~cm is the target length, and 
$A_w$=1.00794~g/mol is the atomic weight of the target.

The statistical uncertainty on the cross section in each bin $i$ includes contributions from the statistical 
uncertainty on the hyperon yield and the acceptance function and is given by:
\begin{equation}
\delta \sigma_i = \sigma_i \left[ \left( \frac{\delta N_i}{N_i} \right)^2 + 
\left( \frac{\delta \eta_i}{\eta_i} \right)^2 \right]^{1/2}.
\end{equation}

\subsection{Particle Identification and Event Selection}
\label{anal-pid}

The $\gamma^* p \to K^+\Lambda$ and $\gamma^* p \to K^+\Sigma^0$ reaction channels were identified by detecting 
a scattered electron in coincidence with a $K^+$ and then using the missing mass technique to identify the 
hyperons. Event reconstruction required the identification of both a final state electron and $K^+$ candidate
within the well-understood fiducial regions of the detector. Details on the algorithms employed to minimize the
particle misidentification at this stage are included in Ref.~\cite{carman_2}. Before computing the missing mass
spectrum, vertex cuts were employed to ensure that the particles originated from the target. In addition, corrections 
to the electron and kaon momenta were devised to account for reconstruction inaccuracies that arose due to  
to relative misalignments of the drift chambers in the CLAS magnetic field, as well as from uncertainties in the 
magnetic field map employed during charged track reconstructions. These corrections were typically less than 1\%.

\begin{figure}[htbp]
\vspace{6.0cm}
\includegraphics{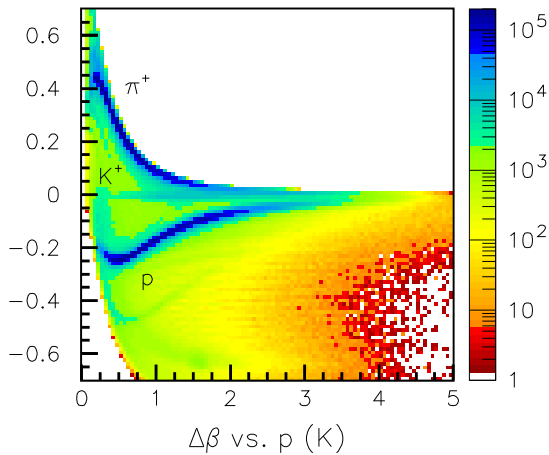}
\caption{(Color online) $\Delta \beta$ vs. momentum (GeV) for the assumption that the reconstructed positively
charged particle was a kaon. The $K^+$ band lies along $\Delta \beta =0$.}
\label{beta-plot}
\end{figure}

The algorithm used for hadron identification relied on comparing the measured velocity $\beta = v/c$ for the track 
candidate to that expected for an assumed $\pi^+$, $K^+$, and $p$ track. The assumption that resulted in the 
minimum $\Delta \beta = \beta - \beta^{calc}_{\pi,K,p}$ was used to identify the species of the track. 
Fig.~\ref{beta-plot} shows $\Delta \beta$ versus momentum for the $K^+$ track assumption. For the data included here, 
the kaon momentum range was between 0.35~GeV (software cut) and $\approx$ 4.5~GeV (kinematic limit), with a typical 
flight path of 5.5~m. The measured mass resolution was primarily due to the reconstructed time-of-flight resolution, 
which was $\approx$100~ps ($\sigma$) on average; it also included contributions from the momentum and path length 
uncertainties of CLAS. Fig.~\ref{beta-plot} shows that unambiguous separation of $K^+$ tracks at the 2$\sigma$ level 
is possible up to about 2~GeV. For higher momenta, the background due to particle misidentification increases. Detailed 
background subtractions are necessary to determine the final event yields.

\begin{figure}[htbp]
\vspace{6.0cm}
\includegraphics{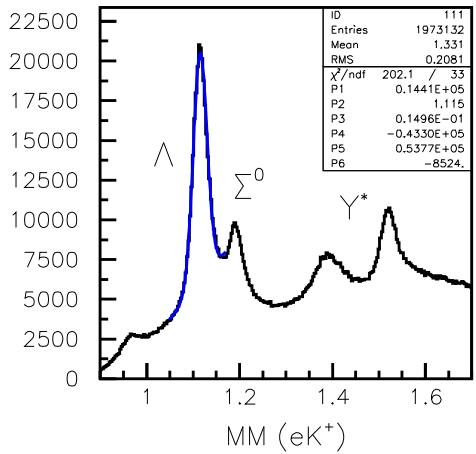}
\caption{(Color online) Distribution of $MM(e'K^+)$ (GeV) showing the $\Lambda$, $\Sigma^0$, and several low-lying
excited hyperon states. These data for the final event sample highlight the hyperon yields relative to the 
underlying background. The fit of the $\Lambda$ peak shows the average mass resolution of $\sigma$=15~MeV for 
this analysis.}
\label{missing-mass}
\end{figure}

Fig.~\ref{missing-mass} shows the $e'K^+$ missing mass ($MM(e'K^+)$) distribution for the final event sample after
all cuts have been made. This distribution contains a background continuum beneath the hyperons that arises due to 
multi-particle final states where the candidate $K^+$ results from a misidentified pion or proton. 

\begin{figure}[htbp]
\vspace{3.3cm}
\includegraphics{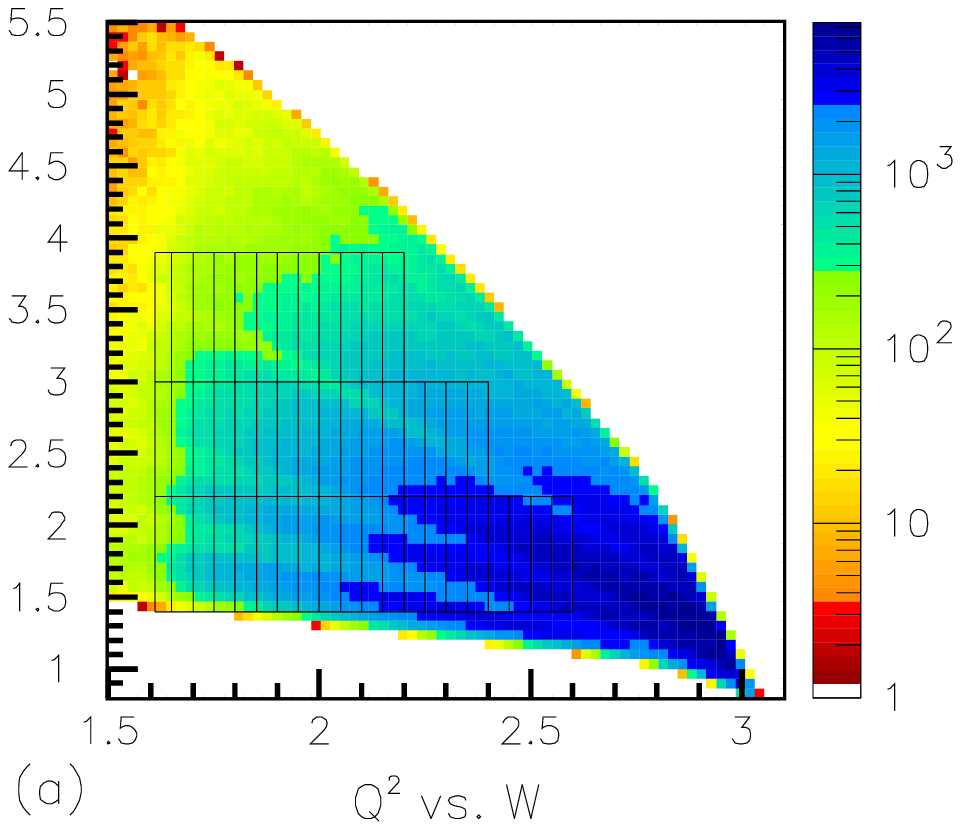}
\includegraphics{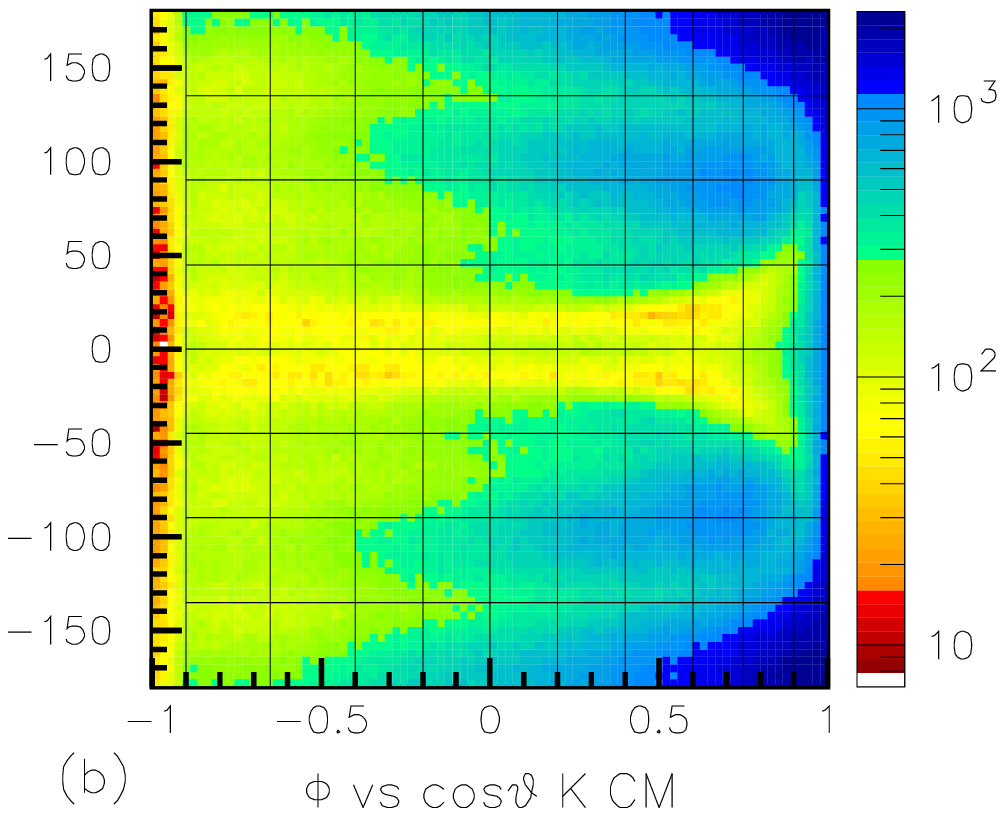}
\caption{(Color online) Kinematic extent of the CLAS e1f data set. (a). $Q^2$ (GeV$^2$) vs. $W$ (GeV). 
(b). $\Phi$ (deg) vs. $\cos \theta_K^*$. The plots are overlaid with the binning choices in this analysis.}
\label{kin}
\end{figure}

The data were binned in a four-dimensional space of the kinematic variables $Q^2$, $W$, $\cos \theta_K^*$,
and $\Phi$. The bin definitions used in this analysis are listed in Table~\ref{tab:bins}. Fig.~\ref{kin}
shows the kinematic extent of the data in terms of $Q^2$ versus $W$ and $\Phi$ versus $\cos \theta_K^*$.
These plots are overlaid with a grid indicating the bins in this analysis. The bin widths in $W$ and $\Phi$ were 
chosen to be uniform. Note that the maximum $W$ bin at each $Q^2$ was limited to where the hyperon yield fits
were not dominated by systematic uncertainties.

\begin{table*}[htbp]
\begin{center}
\begin{tabular}{l|l} \hline
$Q^2$: [1.4,2.2~GeV$^2$]~~ & ~~$W$ : [1.6,2.6~GeV] ~~ (20 50-MeV-wide bins) \\
 ~~~~~ [2.2,3.0~GeV$^2$]~~ & ~~$W$ : [1.6,2.4 GeV] ~~ (16 50-MeV-wide bins) \\
 ~~~~~ [3.0,3.9~GeV$^2$]~~ & ~~$W$ : [1.6,2.2 GeV] ~~ (12 50-MeV-wide bins) \\ \hline
\multicolumn{2} {l} {$\cos \theta_K^*$  : [-0.9,-0.65], [-0.65,-0.4], [-0.4,-0.2], [-0.2,0.0], [0.0,0.2],} \\
\multicolumn{2} {l} {~~~~~~~~~~ [0.2,0.4], [0.4,0.6], [0.6,0.75], [0.75,0.9], [0.9,1.0]} \\ \hline
\multicolumn{2} {l} {$\Phi$: ~~8 bins 45$^\circ$-wide [-180$^\circ$,180$^\circ$]} \\ \hline
\end{tabular}
\end{center}
\caption{Bin limits used for the $KY$ cross sections and structure function analysis in this work.}
\label{tab:bins}
\end{table*}
 
\subsection{Yield Extraction}
\label{anal-yields}

The three components of the $MM(e'K^+)$ spectra are the $K^+\Lambda$ events, the $K^+\Sigma^0$ events, and 
the particle misidentification background (dominated by pions misidentified as kaons). These individual 
contributions must be separated to extract the $K^+\Lambda$ and $K^+\Sigma^0$ differential cross sections 
in each analysis bin.

The approach to separate the signal from the background events employed a fitting process based
on hyperon template shapes and a polynomial to account for the particle misidentification background.
The form for the spectrum fits was given by:
\begin{equation}
MM = A \cdot \Lambda_{template} + B \cdot \Sigma_{template} + P_{bck},
\end{equation}
\noindent
where $\Lambda_{template}$ and $\Sigma_{template}$ are the simulated hyperon distributions with scaling 
factors $A$ and $B$, respectively, and $P_{bck}$ is a polynomial describing the background.

\begin{figure}[htbp] 
\vspace{4.2cm} 
\includegraphics{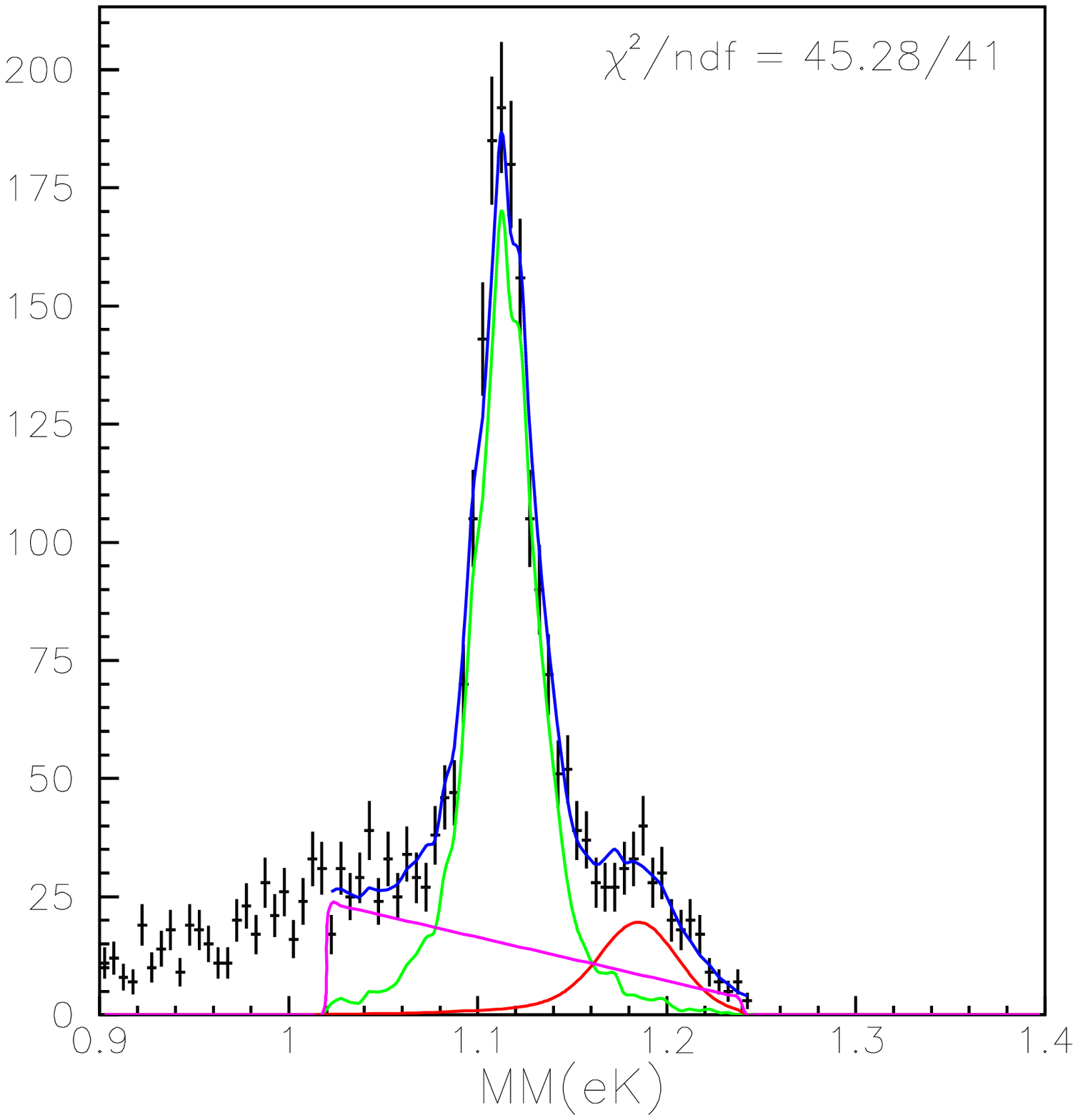}
\includegraphics{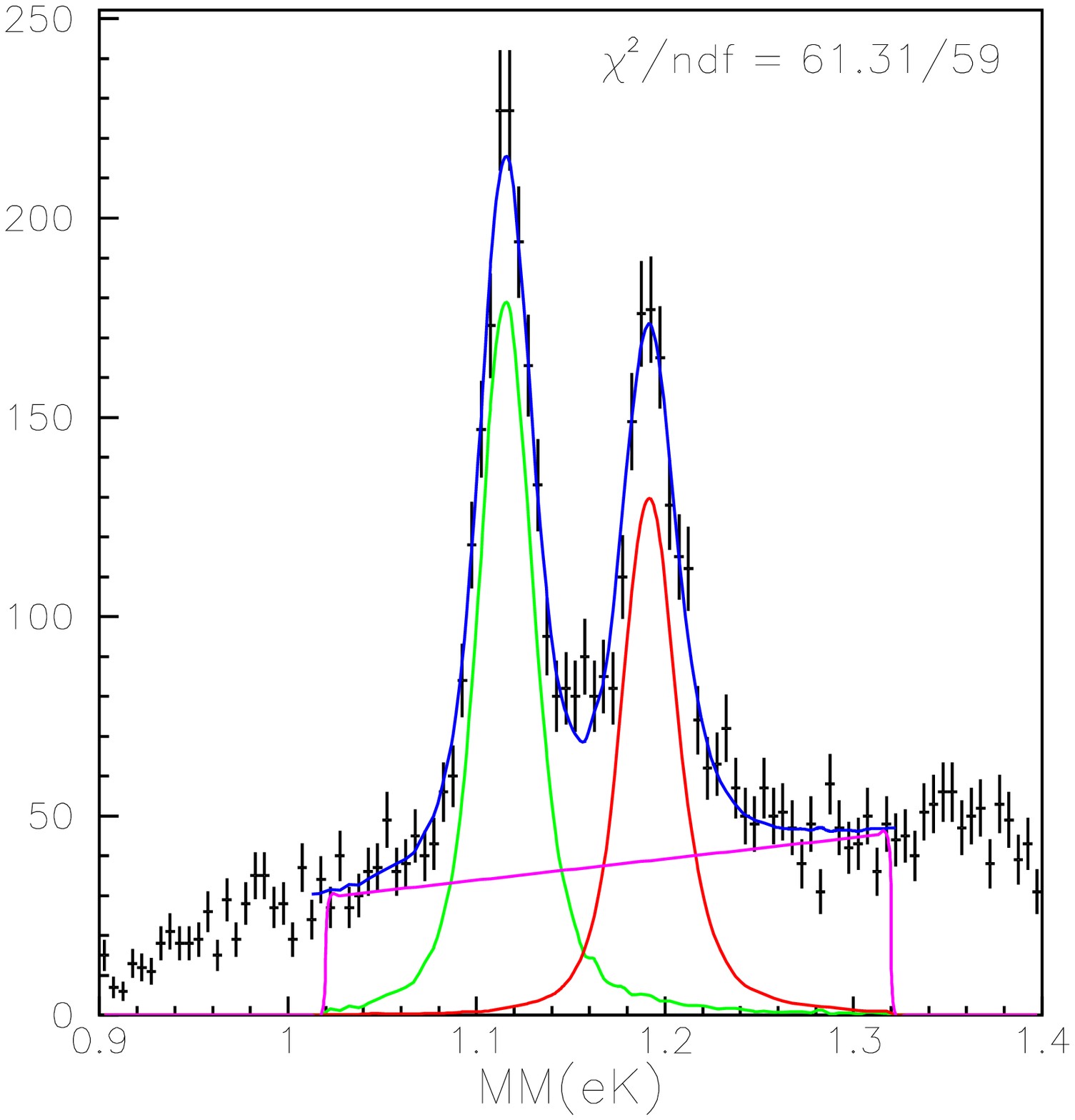}
\caption{(Color online) Sample template fits to the $MM(e'K^+)$ data (GeV) integrated over $Q^2$ and $\Phi$
for $\cos \theta_K^* = 0.10$ and $W$=1.725~GeV (left) and 1.925~GeV (right) to illustrate the typical fit 
quality. The fit includes a $\Lambda$ template, a $\Sigma^0$ template, and a polynomial background term.}
\label{samples}
\end{figure}

The hyperon templates were derived from a GEANT-based Monte Carlo that included radiative processes
and was matched to the detector resolution (see Section~\ref{monte}). The background contributions for 
this fitting were studied with a number of different assumptions (see discussion in Section~\ref{systematics}). 
Ultimately, a linear form for the background was chosen. The template fits to the missing mass spectra 
were carried out using a maximum log likelihood method appropriate for the statistical samples of our data.
Fig.~\ref{samples} shows two sample fits to illustrate the typical fit quality to the data.

The final yields in each kinematic bin were determined by taking the number of counts determined 
from the fits that fell within a mass window around the $\Lambda$ (1.07 to 1.15~GeV) and $\Sigma^0$ 
(1.17 to 1.22~GeV) peaks. Hyperon events in the tails of the distributions that fell outside of 
the mass windows were accounted for by the acceptance and radiative corrections.

The number of $\Lambda$ and $\Sigma^0$ hyperons in both the $K^+\Lambda$ and $K^+\Sigma^0$ mass windows
relative to the total number of counts in the mass windows was found to be independent of $Q^2$ and $\Phi$ 
in each bin of $W$ and $\cos \theta_K^*$. Thus the final yields in each bin were determined by scaling the 
raw yields in the $K^+\Lambda$ and $K^+\Sigma^0$ mass windows by a background factor determined from fits 
in each bin of $W$ and $\cos \theta_K^*$. 

\subsection{Acceptance and Efficiency Corrections}
\label{acc-corr}

\subsubsection{Monte Carlo Acceptance Function}
\label{monte}

Monte Carlo simulations were carried out for this analysis for four distinct purposes. The first was to 
determine the detector acceptance in each bin, the second was as a cross check of the radiative 
correction factors, the third was to generate the hyperon templates for the spectrum fits, 
and the fourth was to determine the tracking efficiency corrections.

\begin{figure}[htbp]
\vspace{6.2cm}
\includegraphics{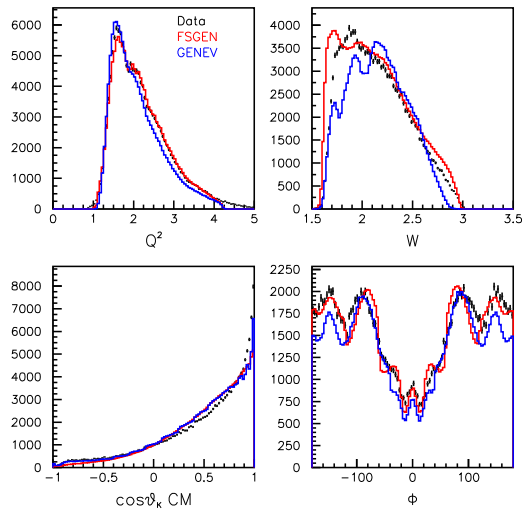}
\caption{(Color online) Comparison between selected $K^+\Lambda$ kinematic distributions ($Q^2$ (GeV$^2$), 
$W$ (GeV), $\cos \theta_K^*$, and $\Phi$ (deg)) of the data (black points with error bars) and the 
corresponding distributions generated from the FSGEN (red - light) and GENEV event generators (blue - dark).}
\label{mc-comp}
\end{figure}

For this analysis we employed two different event generators for the exclusive $K^+\Lambda$ and 
$K^+\Sigma^0$ event samples. The first generator, FSGEN~\cite{fsgen}, generates $ep \to e'K^+Y$ events 
according to a phase space distribution with a $t$-slope scaled by a factor of $e^{-b t}$. This generator 
did not include radiative effects. The nominal choice of the $t$-slope parameter of $b$=1.0~GeV$^{-2}$ was 
chosen to best match the $\cos \theta_K^*$ dependence of the data. The generated data were then weighted
with ad hoc functions so that they matched well to the kinematic distributions of the data (see Fig.~\ref{mc-comp}).

The second generator, GENEV~\cite{genev}, generates events for various meson production channels. It was 
modified for this analysis to include the $K^+\Lambda$ and $K^+\Sigma^0$ channels, reading in cross section 
tables for $K^+\Lambda$ and $K^+\Sigma^0$ photoproduction based on the data of Refs.~\cite{mccracken} and 
\cite{dey}, respectively. It extrapolates to finite $Q^2$ by introducing a virtual photon flux factor and 
electromagnetic form factors based on a simple dipole form. Radiative effects based on the formalism of Mo and Tsai
\cite{mo-tsai} are part of the generator as an option. Here too, the input distributions of the model were 
weighted with ad hoc function so that they matched the data (see Fig.~\ref{mc-comp}).

The Monte Carlo suite is based on a GEANT-3 package~\cite{geant}. The generated events were processed by 
this code based on the CLAS detector. The events were then subjected to additional smearing factors for the 
tracking and timing resolutions to match the average experimental resolutions. The analysis of the Monte Carlo 
data used the same code as was used to analyze the experimental data. Ultimately more than 1 billion Monte Carlo 
events were generated to determine the correction factors and the associated systematic uncertainties, which are 
discussed in Section~\ref{systematics}.

\begin{figure}[htpb]
\vspace{4.2cm}
\includegraphics{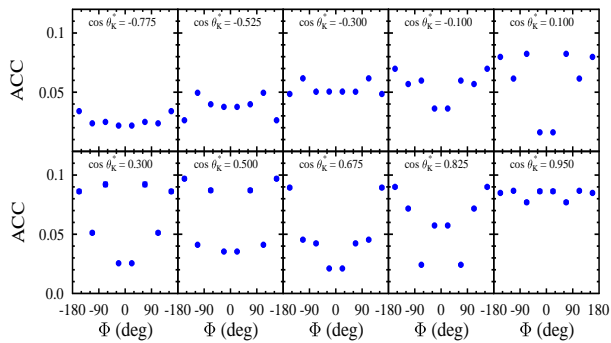}
\caption{Distribution of the computed $K^+\Lambda$ acceptance for CLAS as a function of 
$\cos \theta_K^*$ and $\Phi$ for the $W$=1.925~GeV and $Q^2$=1.8~GeV$^2$ bin. The substructure in
the acceptance is to due to the geometry of the active areas of the CLAS detector. The statistical error 
bars from the Monte Carlo are smaller than the symbol size on this plot.} 
\label{accep}
\end{figure}

In order to relate the experimental yields to the cross sections, we require the detector acceptance to
account for various effects, such as the geometric coverage of the detector, hardware and software 
inefficiencies, and resolution effects from the track reconstruction. The acceptance is defined separately 
for the $K^+\Lambda$ and $K^+\Sigma^0$ reaction channels as a function of the kinematic variables as:
\begin{equation}
Acc_i(Q^2,W,\cos \theta_K^*,\Phi) = \frac{N^{rec}_i(Q^2,W,\cos \theta_K^*,\Phi)}
{N^{gen}_i(Q^2,W,\cos \theta_K^*,\Phi)},
\end{equation}
\noindent
where $N^{rec}_i$ is the reconstructed number of events in each bin and $N^{gen}_i$ is the generated number 
of events in each bin. The FSGEN simulation was used to determine the acceptance function for the final
analysis. Typical acceptances for CLAS for the $e'K^+$ final state vary from $\approx$1\% to 30\%. 
Fig.~\ref{accep} shows examples of this computed acceptance for the $K^+\Lambda$ final state as a function
of $\Phi$ and $\cos \theta_K^*$ for one $Q^2$ and $W$ bin.

\subsubsection{Efficiency Corrections}
\label{eff-corr}

For this analysis several standard CLAS efficiency corrections were applied to the yields on an
event-by-event basis. The first correction accounted for the efficiency of the Cherenkov counter
for registering electron tracks based on the number of detected photoelectrons in each sector in
a fine grid of the $\theta$ and $\phi$ angles of the electron at the face of each CC detector.
The average CC efficiency within the electron geometric fiducial cuts for this analysis is 96\%.

The remaining efficiency corrections account for hadron tracking inefficiencies. The first correction
accounts for the single track reconstruction efficiency in CLAS that is not 100\% due to inefficient
SC paddles and DC tracking regions. This efficiency function was assigned based on the relative ratio
of data counts to Monte Carlo counts as a function of CLAS sector and SC paddle number. These corrections
are at the level of about 10\% on average. 

Another efficiency correction related to tracking is necessary for events in which two charged tracks of
the same charge and similar momenta lie very close to each other. For such events the tracking algorithm
may not successfully identify two separate tracks. For this analysis, a correction was applied to the
small fraction of events in which the $K^+$ and $p$ from the decay of the $\Lambda$ were in the same CLAS sector 
within 10$^\circ$ of each other in polar angle. This efficiency factor is necessary even for the $e'K^+$ 
analysis due the presence of the decay protons in the final state. The systematics associated with each of
these efficiency corrections are discussed in Section~\ref{sys-acc}.

\subsection{Radiative Corrections}
\label{rad-corr}

Radiative effects must be considered when determining the $\gamma^* p \to K^+Y$ cross sections. Radiative
effects result in bin migration such that the measured $Q^2$ and $W$ are not the true $Q^2$ and $W$ to which 
the event should be properly associated.

For this analysis, two different approaches to determine these correction factors have been employed. The 
first uses the stand-alone program EXCLURAD~\cite{exclurad} and the second uses the event generator GENEV
\cite{genev} in combination with the CLAS Monte Carlo. The radiative correction factor that multiplies 
the measured bin-averaged differential cross section in each bin is defined as the ratio of the computed 
bin-averaged cross section with radiation off to that with radiation on. More details on each program are 
included below.

\subsubsection{EXCLURAD}

EXCLURAD represents a covariant technique of cancellation of the infrared divergence that 
leads to independence of any parameter that splits the soft and hard regions of phase space of the radiated 
photons. It uses an integration technique that is exact over the bremsstrahlung photon 
phase space, and thus does not rely on the peaking approximation~\cite{peaking}. This approach is an exact 
calculation in that it specifically accounts for the exclusive nature of the reactions as the detection of 
hadrons in the final state, in addition to the electron, reduces the phase space allowed for the final 
radiative photons. 

The program EXCLURAD was based on the measured structure functions from this analysis for $K^+\Lambda$ and 
$K^+\Sigma^0$. The structure functions $\sigma_U$, $\sigma_{LT}$, $\sigma_{TT}$, and $\sigma_{LT'}$ were 
read into the program and the cross section ratio for each bin in $Q^2$, $W$, $\cos \theta_K^*$, and $\Phi$ 
was computed with radiation off to that with radiation on, giving the radiative correction factor 
$R_i$ for that bin. 

\begin{figure}[htbp]
\vspace{7.2cm}
\includegraphics{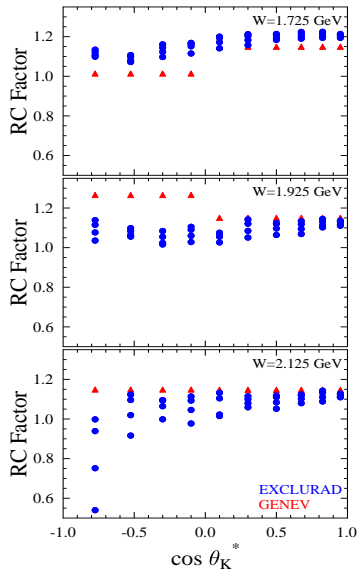}
\caption{(Color online) Radiative corrections factors for the $K^+\Lambda$ final state from EXCLURAD (blue 
- circles) and GENEV (red - triangles) for $Q^2$=1.80~GeV$^2$ as a function of $\cos \theta_K^*$ for representative
values of $W$ of 1.725, 1.925, and 2.125~GeV. The GENEV calculations are independent of $\Phi$ and only one
data point is shown at each value of $\cos \theta_K^*$. The EXCLURAD calculations have a $\Phi$ dependence that is 
symmetric about $\Phi$=0. The different radiative correction values for each $\Phi$ at a given $\cos \theta_K^*$ are 
included on the plot.}
\label{rc-plot1}
\end{figure}

The trends of the correction (shown in Fig.~\ref{rc-plot1}) are such that it has its largest value near threshold
and then quickly falls off to a near constant average value with increasing $W$. Note that the radiative correction 
factors including the helicity-dependent structure function $\sigma_{LT'}$ for the two helicity states have no impact 
on the helicity asymmetry computation in Eq.(\ref{eq:sigltp}) and are not included in the analysis.

\subsubsection{GENEV}
\label{genev-sec}

The event generator GENEV~\cite{genev} was introduced in Section~\ref{monte} as it was used to compute the 
CLAS acceptance function. This program also allows for radiative correction factors to be determined. It 
includes radiative effects based on the formalism for inclusive electron scattering from Ref.~\cite{mo-tsai} 
and employs the peaking approximation~\cite{peaking} in the computation. As GENEV is based on an evolution of 
the photoproduction cross sections, it does not have an explicit $\Phi$ dependence and thus the $R_i$ factors 
in Eq.(\ref{dcs}) were determined in bins of $Q^2$, $W$, and $\cos \theta_K^*$.

This model has several shortcomings. The first is that the phase space for the radiated 
photons is not properly computed as this is modified by the detected hadrons. Secondly, the model is based on 
only the longitudinal and transverse response and does not include the interference structure functions 
$\sigma_{LT}$ or $\sigma_{TT}$. Finally, the approach relies on an unphysical parameter to split the hard and 
soft regions of the radiated photon phase space to cancel the infrared divergence. Due to the known limitations 
with this approach, it was used only to provide a qualitative cross check to the EXCLURAD results and to explore 
the associated systematic uncertainties (see Section~\ref{sys-acc}). Fig.~\ref{rc-plot1} shows a comparison of 
the radiative correction factors computed by GENEV to those computed from EXCLURAD. Apart from the region near 
threshold, the correspondence between the two approaches is within 10\%.

\subsection{Bin Centering Corrections}
\label{bin-center}

The goal of this analysis is to measure cross sections and separated structure functions for the $K^+Y$ 
final states at specific kinematic points. However, the analysis proceeds from using finite bins in the 
relevant kinematic quantities $Q^2$, $W$, $\cos \theta_K^*$, and $\Phi$ (see Section~\ref{anal-pid}). 

The virtual photon flux factor $\Gamma_v$ defined in Section~\ref{formalism} is computed for each bin 
using the bin-averaged values of $Q^2$ and $W$. If the cross sections were computed at this point using 
Eq.(\ref{dcs}) with the $BC_i$ terms set to unity, we would have completed a measurement of the bin-averaged 
cross sections that we could quote at the corresponding bin-averaged kinematic points. To quote the cross 
section at specific kinematic points of our choosing, namely, the geometric centers of the defined bins, we 
must evolve the cross sections from the bin-averaged kinematic points to the geometric bin centers. 
These evolution factors are the bin-centering correction factors $BC_i$ in Eq.(\ref{dcs}).
The bin-centering corrections are then applied for each bin as:
\begin{equation}
\frac{d\sigma}{d\Omega}^{point}_i = \frac{d\sigma}{d\Omega_i}^{avg} 
\left( \frac{\frac{d\sigma}{d\Omega}^{point}}{\frac{d\sigma}{d\Omega}^{avg}} \right )^i_{model} = 
\frac{d\sigma}{d\Omega_i}^{avg} \cdot BC_i,
\end{equation}
\noindent
where $BC_i$ are the ratios of the bin-centered cross section to the bin-averaged cross section.

Studies of the bin-averaged kinematic quantities versus the geometric bin-centered values show that
there is no need for bin-centering corrections in $W$ or $\cos \theta_K^*$. For this work the threshold
$W$ bin for $K^+\Lambda$ is quoted at 1.630~GeV and for $K^+\Sigma^0$ at 1.695~GeV. To determine the 
bin-centering factor $BC_i$ for each bin, we have fit the measured structure functions $\sigma_U$  
for each $W$ and $\cos \theta_K^*$ bin versus $Q^2$ for both the $K^+\Lambda$ and $K^+\Sigma^0$ final 
states. To bin center the data at specific $Q^2$ points, we have used the following dipole evolution factor:
\begin{equation}
\label{bc-eq}
BC_i = \frac{\left( 1 + Q^2_{point}/0.7 \right)^{-2}_i}{\left( 1 + Q^2_{avg.}/0.7 \right)^{-2}_i}
~~~(Q^2~{\rm in~GeV}).
\end{equation}

The bin centering factors using this form were in the range from 0.95 to 1.05 across the full kinematic phase
space.

\subsection{Structure Function Extraction}
\label{sfsep}

The differential cross sections computed using Eq.(\ref{sigma0}) are the mean values within the finite size 
of the $\Phi$ bins and therefore do not reflect the value at the bin center. Thus directly fitting 
these data with Eq.(\ref{sigma0}) to extract the structure functions $\sigma_U = \sigma_T + \epsilon \sigma_L$,
$\sigma_{TT}$, and $\sigma_{LT}$ would be inappropriate.  Integrating Eq.(\ref{sigma0}) over the finite bin size, 
$\Delta\Phi=\Phi_u-\Phi_l$, where $\Phi_u$ and $\Phi_l$ are the upper and lower limits of the bin, 
respectively, gives:
\begin{multline}
\bar{\sigma}_0 \equiv \frac{1}{\Delta\Phi}\int_{\Phi_l}^{\Phi_u}
( \sigma_U + \epsilon\sigma_{TT}\cos 2\Phi + \\ 
\sqrt{\epsilon(\epsilon+1)}\sigma_{LT}\cos\Phi ) d\Phi \\
=\frac{1}{\Delta\Phi} ( \sigma_U \Delta\Phi +
\frac{\epsilon}{2} \sigma_{TT} \left( \sin 2\Phi_u-\sin 2\Phi_l \right) + \\
\sqrt{\epsilon(\epsilon+1)} \sigma_{LT} \left( \sin\Phi_u-\sin\Phi_l \right) ).
\label{eq-csecfit}
\end{multline}
\noindent
$\bar{\sigma}_0$ now represents the value of the measured bin-averaged cross section in a given $\Phi$ bin and 
fitting the data with Eq.(\ref{eq-csecfit}) yields the separated structure functions.  The ``$\epsilon$'' 
pre-factors were evaluated at the bin center and divided out. Note that prior to the $\Phi$ fits, the statistical 
uncertainty on each cross section point was combined linearly with that portion of the systematic uncertainty 
arising from the yield extraction procedures (see Section~\ref{yield-sys} for details). 

\begin{figure}[htbp]
\vspace{6.7cm}
\includegraphics{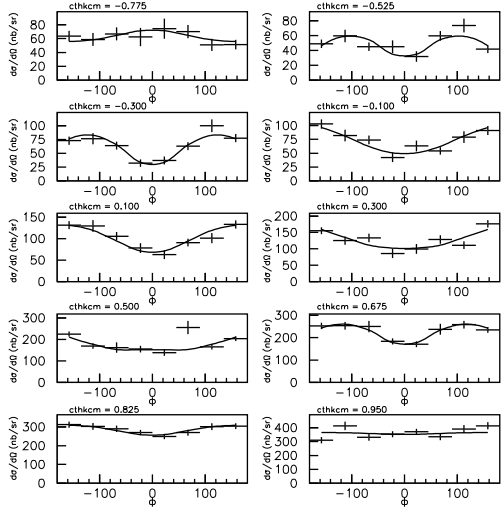}
\caption{Fits of the $K^+\Lambda$ differential cross section (nb/sr) vs. $\Phi$ (deg) for $W$=1.725~GeV and 
$Q^2$=1.8~GeV$^2$ showing a sample of the fits to extract $\sigma_U$, $\sigma_{LT}$, and 
$\sigma_{TT}$.}
\label{phi-fits1}
\end{figure}

In Fig.~\ref{phi-fits1} we show a sample of the $\Phi$-dependent differential cross sections for the 
$K^+\Lambda$ final state at $W$=1.725~GeV for $Q^2$=1.8~GeV$^2$. The different shapes of the differential 
cross sections versus $\Phi$ in each of our bins in $Q^2$, $W$, and $\cos \theta_K^*$ reflect differences 
of the interference terms, $\sigma_{LT}$ and $\sigma_{TT}$, while the differences in scale reflect the 
differences in $\sigma_U$.

The extraction of $\sigma_{LT'}$ in each bin of $Q^2$, $W$, and $\cos \theta_K^*$ requires knowledge of both 
the asymmetry $A_{LT'}$ and the unpolarized cross section $\sigma_0$, which can be seen by rearranging
Eq.(\ref{eq:sigltp}) into a normalized asymmetry $A_N^{meas}$ as:
\begin{equation}
\label{norma}
A_N^{meas} = \frac{A_{LT'}\sigma_0}{\sqrt{\epsilon (1 - \epsilon)}} = \sigma_{LT'} \sin \Phi.
\end{equation}

$A_{LT'}$ is determined by forming the asymmetry of the $K^+\Lambda$ and $K^+\Sigma^0$ yields for the positive
and negative beam helicity states ($h=\pm 1$) as:
\begin{equation}
A_{LT'} = \frac{1}{P_b} \left( \frac{N^+ - N^-}{N^+ + N^-} \right ),
\end{equation}
\noindent
where $P_b$ is the average longitudinal polarization of the electron beam.

\begin{figure}
\vspace{6.7cm} 
\includegraphics{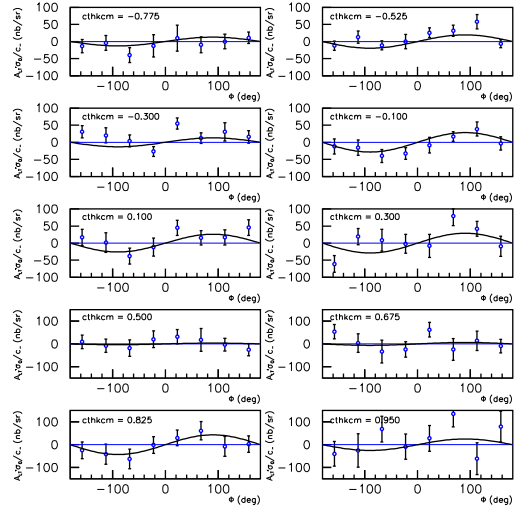}
\caption{Normalized asymmetries for $K^+\Lambda$ vs. $\Phi$ (deg) at for $W$=1.725~GeV and $Q^2$=1.8~GeV$^2$
showing a sample of the fits to extract $\sigma_{LT'}$.}
\label{phi-fits2}
\end{figure}

As with the cross sections, the measured asymmetries are the average values over the span of the given $\Phi$ 
bins. Integrating Eq.(\ref{eq:sigltp}) over the size of the $\Phi$ bin results in:
\begin{equation}
\label{normb}
A_N = A_N^{meas} \frac{\sin \Phi \Delta \Phi}{\cos \Phi_l - \cos \Phi_u}.
\end{equation}
To extract $\sigma_{LT'}$, a $\sin \Phi$ fit was performed according to Eq.(\ref{normb}), where the
kinematic $\epsilon$ factor was calculated at the bin-centered values of $Q^2$ and $W$ for each bin.
A sample of these distributions is shown in Fig.~\ref{phi-fits2} for the $K^+\Lambda$ final state at
$W$=1.725~GeV for $Q^2$=1.8~GeV$^2$. Similar to the case for the unpolarized structure function extraction
discussed in Section~\ref{sfsep}, prior to the $\Phi$ fits the statistical uncertainty on the helicity-gated
yields was combined linearly with that portion of the systematic uncertainty arising from the yield
extraction procedure (see Section~\ref{yield-sys} for details).

The statistical uncertainty on the data points in each bin $i$ are a combination of the contributions from both 
$A_{LT'}$ and $\sigma_0$ and are given by:
\begin{equation}
\delta (A_{LT'} \sigma_0)_i = \sqrt{ (A_{LT'} \delta \sigma_0)^2_i + (\sigma_0 \delta A_{LT'})^2_i}.
\end{equation}

\section{Systematic Uncertainties}
\label{systematics}

To obtain a virtual photoabsorption cross section, we extract the yields for the $K^+\Lambda$ and $K^+\Sigma^0$ 
reactions from the the missing-mass spectra for each of our bins in $Q^2$, $W$, $\cos \theta_K^*$, and $\Phi$. 
The yields are corrected for the acceptance function of CLAS including various efficiency factors, radiative effects, 
and bin-centering factors. Finally, we divide by the virtual photon flux factor, the bin volume corrected for 
kinematic limits, and the beam-target luminosity to yield the cross section. Each of these procedures 
is subject to systematic uncertainty. We typically estimate the size of the systematic uncertainties by 
repeating a procedure in a slightly different way, e.g. by varying a cut parameter within reasonable 
limits, by employing an alternative algorithm, or by using a different model to extract a correction, and noting 
how the results change. 

\begin{table*}[htbp]
\begin{center}
\begin{tabular} {|l|c|c|c|c|} \hline
~~~~~~~~Category              & $d\sigma/d\Omega$ & $\sigma_U$ & $\sigma_{LT}$,~$\sigma_{TT}$ &  $\sigma_{LT'}$ \\ \hline \hline
1. Yield Extraction           & \multicolumn{4} {|c|} { }      \\ \hline
~~~~~Signal fitting/binning effects~ & \multicolumn{4} {|c|} {$1.20 \times$~stat.err.} \\ \hline
~~~~~Fiducial cuts                  & 0.4-2.6\% & 0.4-2.6\%  & -             & 0.7-4.4\% \\ \hline
~~~~~Electron identification        & 1.1\%     & 0.1\%      & 4.0\%         & 1.4\% \\ \hline
2. Detector Acceptance        & \multicolumn{4} {|c|} { }      \\ \hline 
~~~~~MC model dependence      & 4.0-9.3\%        & 3.6-7.8\%  & 6.8\%        & 3.6-7.0\% \\ \hline
~~~~~Tracking efficiencies    & 5.3\%            & 5.3\%      & 5.5\%        & 5.3\% \\ \hline
~~~~~Close track efficiencies & 2.8\%             & 1.6\%      & 4.7\%       & 2.6\% \\ \hline
~~~~~CC efficiency function   & 1.5\%             & 1.5\%      & 1.5\%       & 1.5\% \\ \hline
3. Radiative Corrections      & 2.0\%             & 2.0\%      & 4.4\%       & 2.0\% \\ \hline
4. Bin Centering              & 0.5\%             & 0.5\%      & 0.5\%       & 0.5\% \\ \hline
5. Scale Uncertainties        & \multicolumn{4} {|c|} { }      \\ \hline
~~~~~Beam polarization       & -                 & -          & -            & 2.3\% \\ \hline
~~~~~Photon flux factor      & 3.0\%             & 3.0\%      & -            & 3.0\% \\ \hline
~~~~~Luminosity              & 3.0\%             & 3.0\%      & -            & 3.0\% \\ \hline \hline 
Total $Q^2_1$                & 12.5\%            & 11.1\%     & 11.7\%       & 11.6\% \\ \hline
Total $Q^2_2$                &  9.2\%            &  8.2\%     & 11.7\%       & 9.2\% \\ \hline
Total $Q^2_3$                &  8.9\%            &  8.5\%     & 11.7\%       & 9.0\% \\ \hline
\end{tabular}
\caption{Categories and systematic uncertainty assignment for the observables reported in this work for
our three $Q^2$ points at $Q^2_1$=1.80, $Q^2_2$=2.60, and $Q^2_3$=3.45~GeV$^2$. The total systematic uncertainty
assignments for each $Q^2$ point are obtained by adding the different contributions in quadrature.}
\label{syserror}
\end{center}
\end{table*}

In this section we describe our main sources of systematics. The five categories of systematic uncertainty 
studied in this analysis include yield extraction, detector acceptance, radiative corrections, bin centering 
corrections, and scale uncertainties. Each of these categories is explained in more detail below.

In assigning the associated systematic uncertainties, we have compared the differential cross sections and 
extracted structure functions, $\sigma_U$, $\sigma_{LT}$, $\sigma_{TT}$, and $\sigma_{LT'}$, with 
the nominal cuts and the altered cuts. The fractional uncertainty for each bin $i$ was calculated via:
\begin{equation} 
\label{sys1-diff}
\delta\sigma_i  = \frac{\sigma_i^{nom} - \sigma_i^{mod}}{\sigma_i^{nom}}.
\end{equation}

The relative difference in the results $\delta \sigma_i$ is then used as a measure of the systematic uncertainty. 
In this analysis we have carefully studied the kinematic dependence of the systematics and conclude that 
there is no evidence within a given $Q^2$ bin of systematic variations with $W$, $\cos \theta_K^*$, or 
$\Phi$. Table~\ref{syserror} lists the categories, specific sources, and the assigned systematic uncertainties 
on our measurements. Overall the scale of the systematic uncertainties is at the level of about 10\%. 

\subsection{Yield Extraction}
\label{yield-sys}

The procedure to determine the $K^+Y$ yields in each analysis bin employs hyperon templates derived from 
Monte Carlo simulations that have been tuned to match the data. The background fit function has 
been studied using two different approaches. The first uses a polynomial (either linear or quadratic) 
and the second uses the $ep \to e'\pi^+X$ data sample purposefully misidentifying the detected $\pi^+$ 
as a $K^+$. We have concluded that all systematic effects associated with the spectrum fitting get
larger in direct proportion to the size of the statistical uncertainty. We estimated that the systematic 
uncertainty due to the yield extraction is roughly equal to 20\% of the size of the statistical uncertainty 
in any given bin. We added these correlated uncertainties linearly with the statistical uncertainties on 
our extracted yields before performing the $\Phi$ fits.

The other sources of systematic uncertainty considered in this category are associated with the defined 
electron and hadron fiducial cuts and the cuts on the deposited energy in the calorimeter used to identify 
the candidate electron sample. Variations in the definitions of the fiducial cuts and the EC energy
cuts over a broad range showed that the observables were stable for each cut type to within 5\%.

\subsection{Detector Acceptance}
\label{sys-acc}

In the category of detector acceptance, the associated systematics include that due to the model
dependence of the acceptance function, the stability of the tracking efficiency corrections, and
the CC efficiency function.

For this analysis both the FSGEN and GENEV physics models were used to generate the Monte Carlo events. 
Because of the finite bin sizes used in this analysis, it is necessary to study how the derived acceptance 
function based on the different event generators impacts the extracted observables. For both models we 
determined the acceptance function and stepped through the full analysis chain to extract the observables.
The systematics assigned for the model dependence were in the range from about 4\% to 9\%.

The approach to assign a systematic associated with the CLAS tracking efficiency corrections was to employ 
slightly different algorithms and then to step through the full analysis chain. The tracking efficiency gave 
stable results at the level of 5\%. The systematic associated with the close track efficiency was stable in 
the range from 2 to 5\%.

To study the systematic uncertainty associated with the CC efficiency function, we compared the measured
observables with the nominal CC efficiency corrections to an analysis with the CC efficiency set to 100\% 
for all events. The differences were within 1.5\% for all observables.

\subsection{Radiative Corrections}

Two very different approaches have been used to study the radiative corrections for the $K^+\Lambda$ and 
$K^+\Sigma^0$ electroproduction reactions. The first was the exclusive approach based on the EXCLURAD 
program~\cite{exclurad} and the second was based on the inclusive approach based on the GENEV program
\cite{genev}. Comparison of the extracted radiative corrections between EXCLURAD and GENEV were within 
about 8\% of each other. However, due to the shortcomings of the GENEV model as discussed in 
Section~\ref{genev-sec}, this comparison was only used as a cross check of the overall scale of the 
corrections.

To assign a systematic uncertainty for the radiative corrections for this analysis, we compared the
measured observables using the EXCLURAD approach but varying the energy range of integration of the
radiated photon over a broad range. The corrections were stable in the range from 2 to 5\%. 

\subsection{Bin Centering Corrections}

To assign a systematic uncertainty to the bin centering corrections, the mass term in the dipole form 
(see Eq.(\ref{bc-eq})) was varied over a broad range. The maximum variation seen in any of the extracted 
observables was 0.5\%.

\subsection{Scale Uncertainties}

In the category of scale uncertainties, the associated systematics include that due to the beam-charge
asymmetry and uncertainties in the beam polarization, the photon flux factor, and the luminosity.

The estimated beam-charge asymmetry is at the level of a few times $10^{-4}$ and is thus entirely 
negligible. The uncertainty in the beam polarization affects only the systematic assigned to $\sigma_{LT'}$.
This is given by: 
\begin{equation}
\delta \sigma_{LT'} = |A_{LT'}^{meas}| \frac{\delta P_e}{P_e} = |\sigma_{LT'}| 0.023,
\end{equation}
\noindent
where $\delta P_e$=0.03 and $P_e$ = 0.754 is the average beam polarization. Thus the assigned systematic 
for $\sigma_{LT'}$ due to the beam polarization uncertainty is 2.3\%.

The uncertainties in the average virtual photon flux factor across our phase space were estimated by 
propagating through the flux definition the uncertainties associated with $W$ and $Q^2$ that arise 
from the uncertainty in the reconstructed electron momentum and angles. The uncertainty in the flux factor 
was determined to be 3\%. This scale-type uncertainty affects only the differential cross section and the 
structure functions $\sigma_U$ and $\sigma_{LT'}$.

We estimated uncertainties in the beam-target luminosity based on the analysis of CLAS $ep$ elastic 
scattering cross sections from Ref.~\cite{elastics}. The overall systematic uncertainty of the Faraday Cup 
charge measurement has been assigned to be 3.0\%. This scale-type uncertainty affects only the differential 
cross section and the structure functions $\sigma_U$ and $\sigma_{LT'}$.

\subsection{Cross Checks}
\label{cross-check}

The nominal analysis for the $K^+\Lambda$ and $K^+\Sigma^0$ differential cross sections and separated structure 
functions required only the detection of the electron and $K^+$ in the final state. In order to check the overall 
systematic assignment, the observables were also extracted when detecting an additional $p$. The detection of the 
proton from the $\Lambda$ decay gives rise to an analysis sensitive to the same systematic uncertainties as the 
nominal analysis, and thus should yield consistent results. However, requiring the proton reduces the acceptance 
by roughly a factor of three, therefore this comparison can only be used as a cross check of the nominal analysis.

The agreement between the cross sections extracted using the $e'K^+$ and $e'K^+p$ final states is  
at the level of $\pm$5-10\% and independent of kinematics to within the statistical uncertainties. 
The differences are driven by the marginal statistics in some of the analysis bins for the $e'K^+p$ 
analysis. These comparisons show that the assigned systematic uncertainties are reasonable.

\section{Results and Discussion}
\label{results}

\subsection{Angular Dependence}

\begin{sidewaysfigure*}[htbp]
\vspace{14.0cm}
\includegraphics{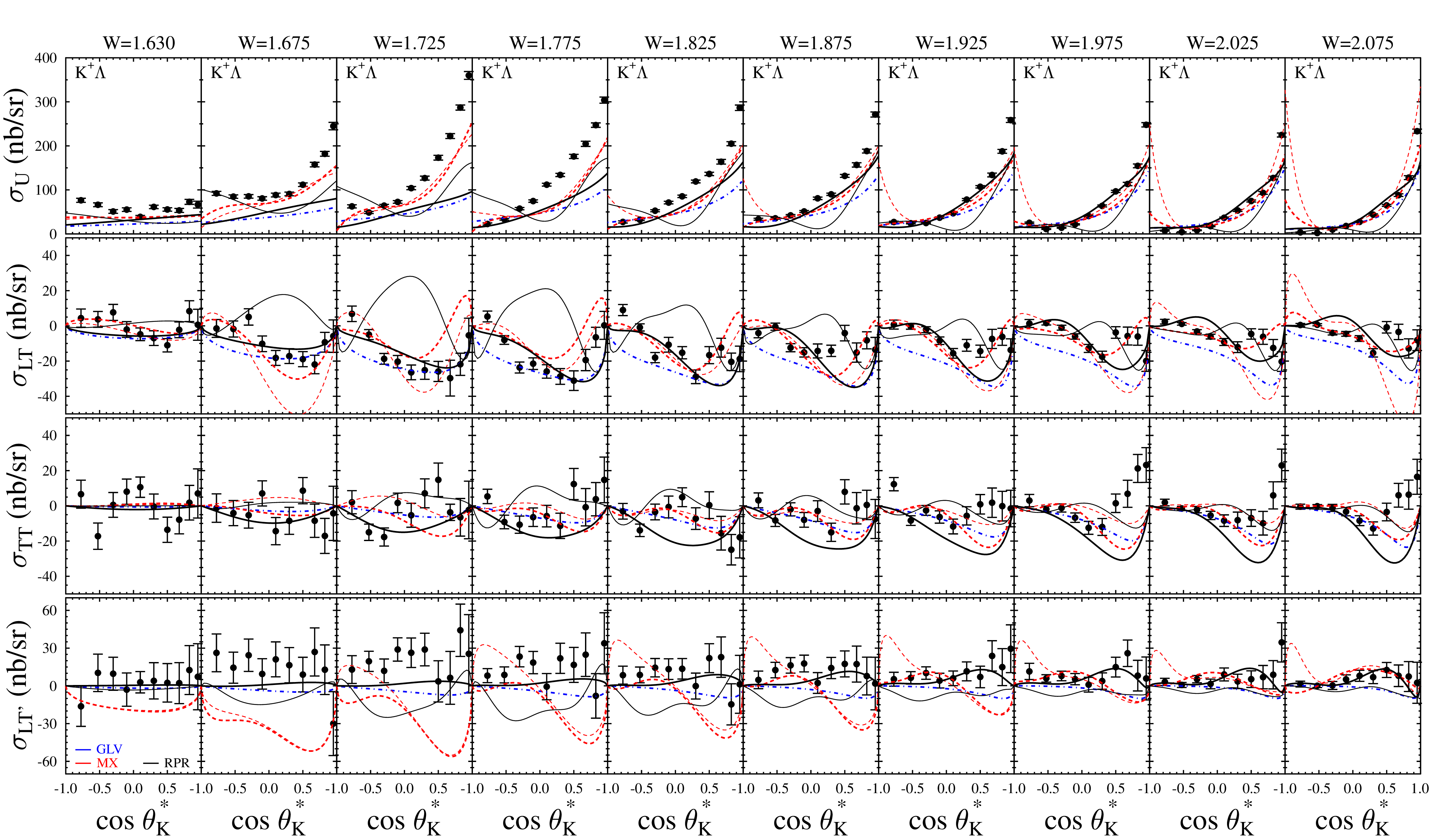}
\caption{(Color online) Structure functions $\sigma_U$, $\sigma_{LT}$, $\sigma_{TT}$, and $\sigma_{LT'}$ 
(in nb/sr) for $K^+\Lambda$ production vs. $\cos \theta_K^*$ at 5.499~GeV for $Q^2$=1.80~GeV$^2$ and $W$ 
from 1.630 to 2.075~GeV. The error bars represent the statistical uncertainties only. The curves shown 
are from the model calculations of Maxwell {\it et al.} (MX) (red/dashed curves)~\cite{maxwell1,maxwell2,maxwell3}, 
Guidal {\it et al.} (GLV) (blue/dot-dashed curves)~\cite{glv}, and Ghent (RPR) (black/solid curves)~\cite{corthals}. 
See the text for detailed descriptions of the calculations and the corresponding references.}
\label{lam_q1_ca} 
\end{sidewaysfigure*}

\begin{sidewaysfigure*}[htbp]
\vspace{14.0cm}
\includegraphics{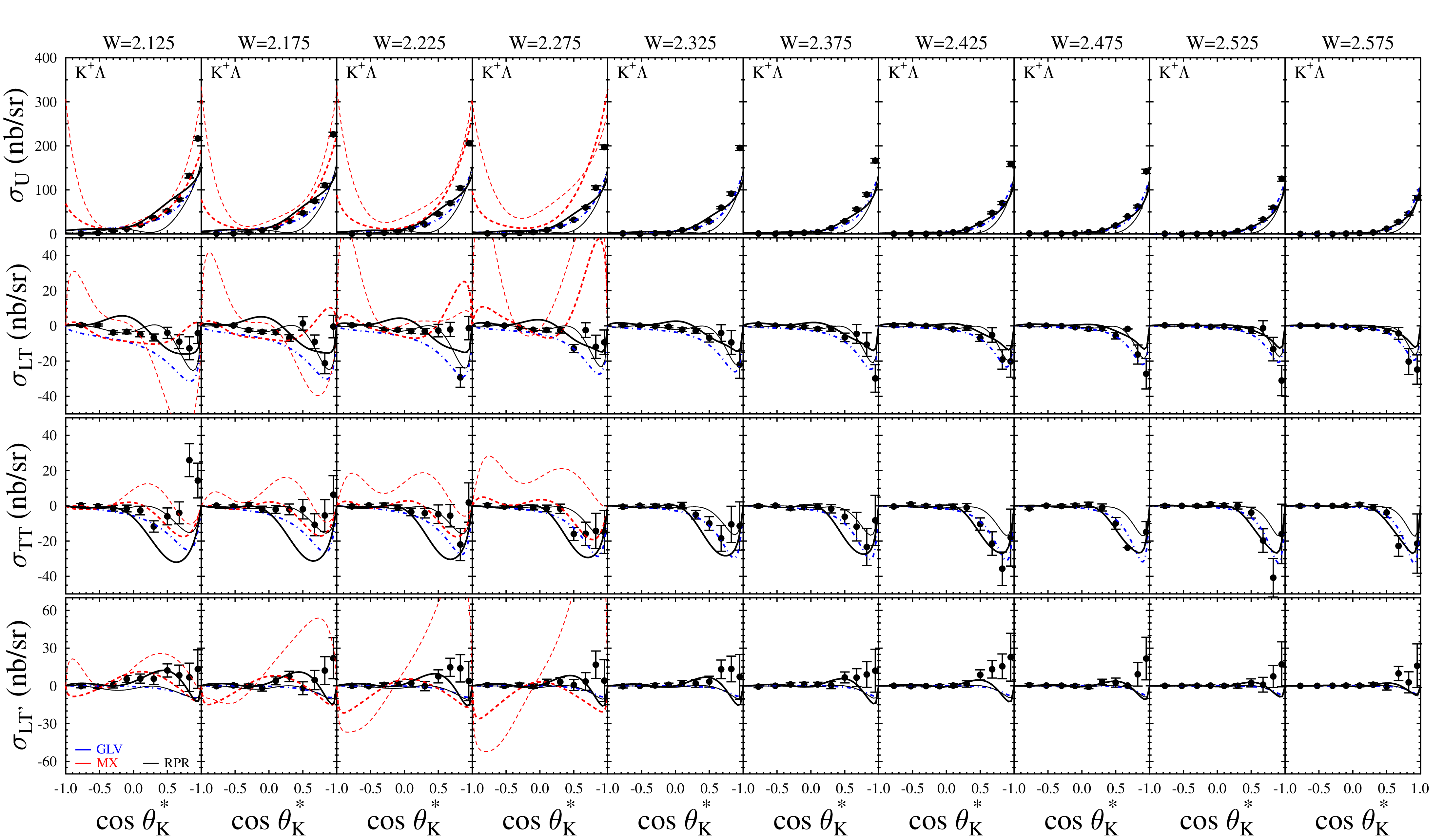}
\caption{(Color online) Structure functions $\sigma_U$, $\sigma_{LT}$, $\sigma_{TT}$, and $\sigma_{LT'}$ 
(in nb/sr) for $K^+\Lambda$ production vs. $\cos \theta_K^*$ at 5.499~GeV for $Q^2$=1.80~GeV$^2$ and $W$ 
from 2.125 to 2.575~GeV. The error bars represent the statistical uncertainties only. The curves are
defined in the caption of Fig.~\ref{lam_q1_ca}.}
\label{lam_q1_cb} 
\end{sidewaysfigure*}

\begin{sidewaysfigure*}[htbp]
\vspace{14.0cm}
\includegraphics{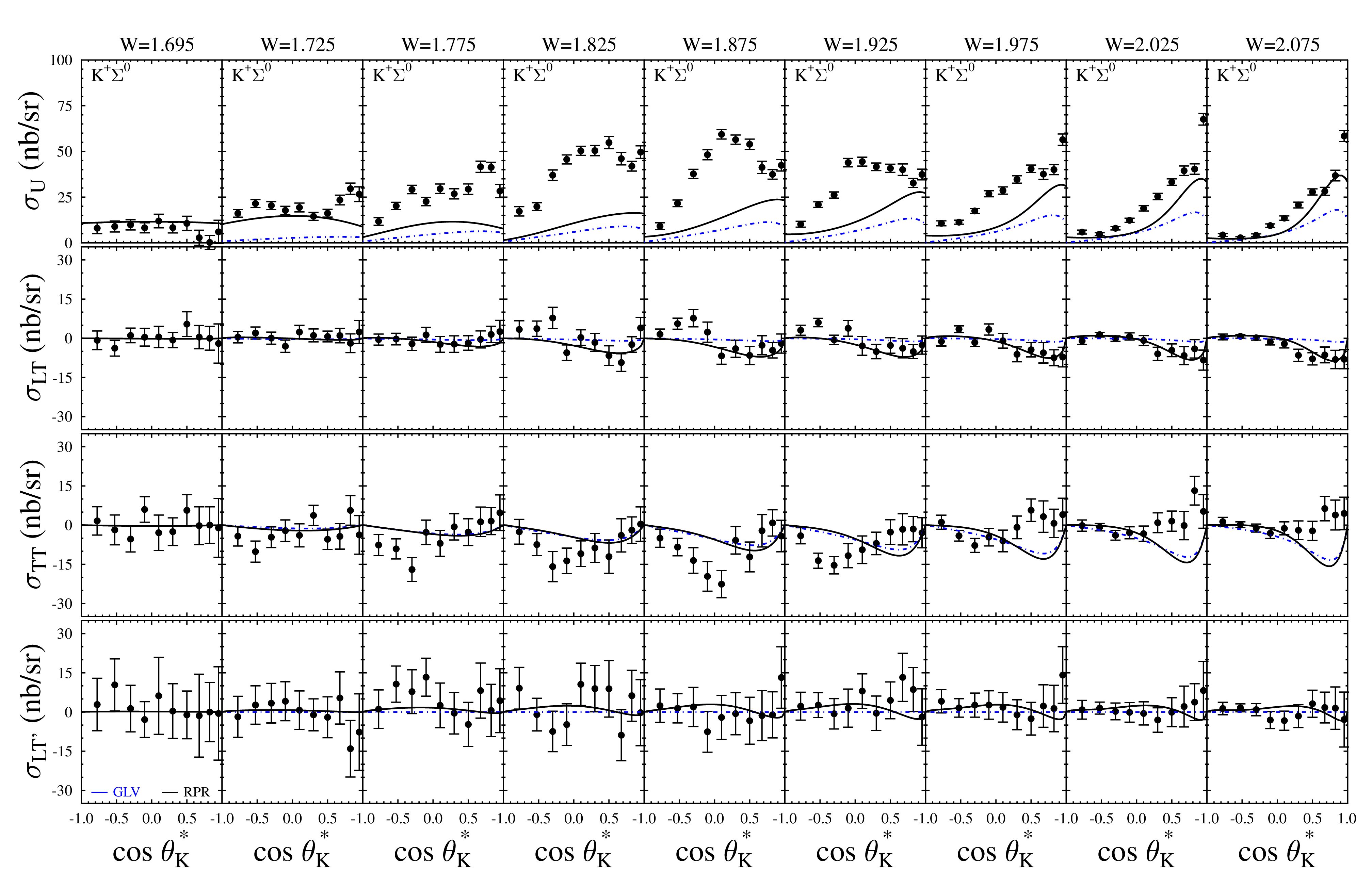}
\caption{(Color online) Structure functions $\sigma_U$, $\sigma_{LT}$, $\sigma_{TT}$, and $\sigma_{LT'}$ 
(in nb/sr) for $K^+\Sigma^0$ production vs. $\cos \theta_K^*$ at 5.499~GeV for $Q^2$=1.80~GeV$^2$ and $W$ 
from 1.630 to 2.075~GeV. The error bars represent the statistical uncertainties only. The curves are
defined in the caption of Fig.~\ref{lam_q1_ca}.}
\label{sig_q1_ca} 
\end{sidewaysfigure*}

\begin{sidewaysfigure*}[htbp]
\vspace{14.0cm}
\includegraphics{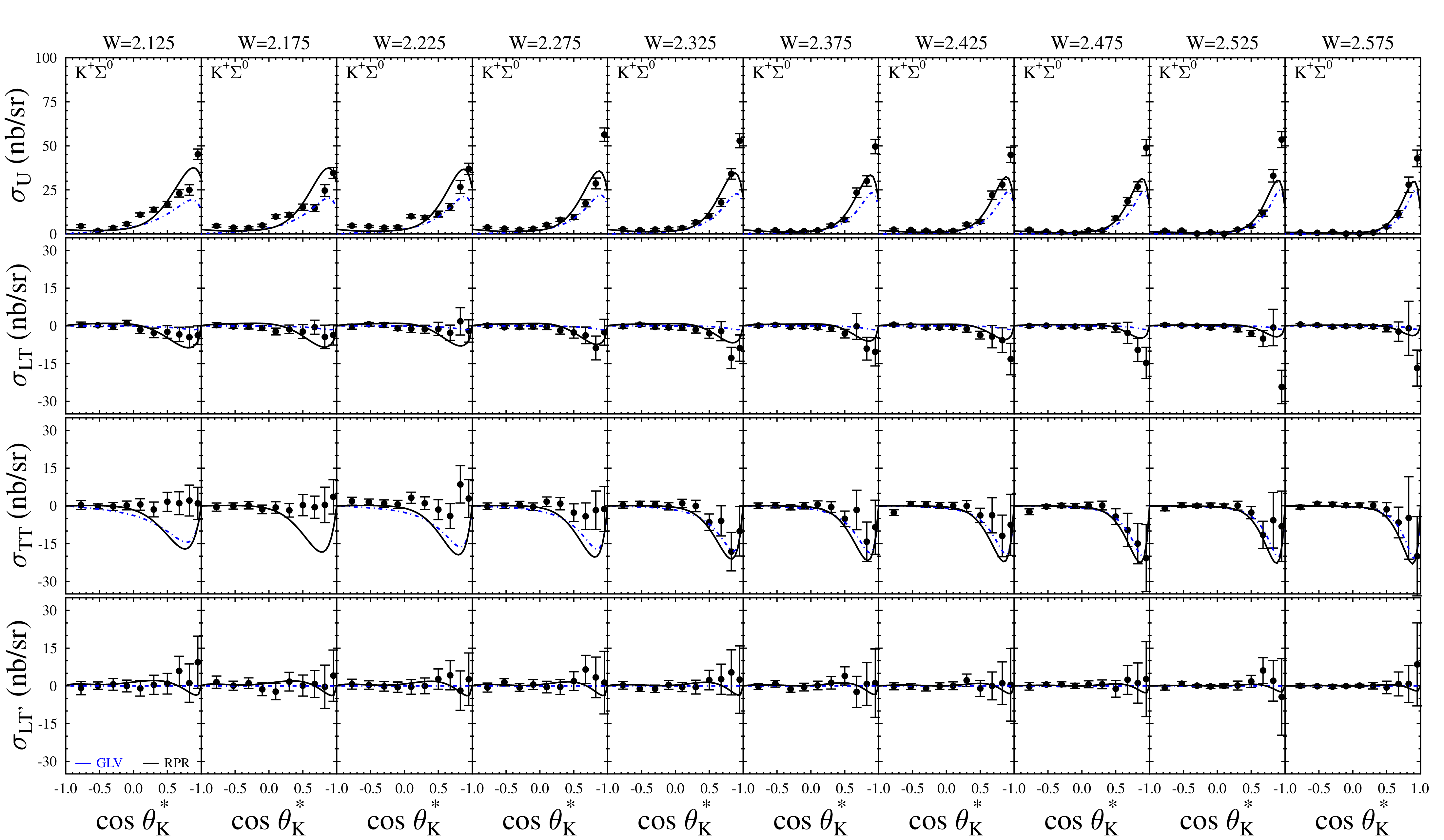}
\caption{(Color online) Structure functions $\sigma_U$, $\sigma_{LT}$, $\sigma_{TT}$, and $\sigma_{LT'}$ 
(in nb/sr) for $K^+\Sigma^0$ production vs. $\cos \theta_K^*$ at 5.499~GeV for $Q^2$=1.80~GeV$^2$ and $W$ 
from 2.125 to 2.575~GeV. The error bars represent the statistical uncertainties only. The curves are 
defined in the caption of Fig.~\ref{lam_q1_ca}.}
\label{sig_q1_cb} 
\end{sidewaysfigure*}

In Figs.~\ref{lam_q1_ca} and \ref{lam_q1_cb} we show the extracted structure functions $\sigma_U$,
$\sigma_{LT}$, $\sigma_{TT}$, and $\sigma_{LT'}$ versus $\cos \theta_K^*$ for the $K^+\Lambda$
final state. Figs.~\ref{sig_q1_ca} and \ref{sig_q1_cb} show the same plots for the $K^+\Sigma^0$
final state. These plots are for our lowest $Q^2$ point at 1.80~GeV$^2$. The general conclusions that 
can be drawn from studying the angular dependence are similar for the two higher $Q^2$ points at 2.60 
and 3.45~GeV$^2$. However, the full set of our data is available in the CLAS physics database~\cite{database}.

The following curves are overlaid on the data:

\begin{itemize}
\item The hadrodynamic model of Maxwell {\it et al.} (MX) (red/dashed curves - thinner 
line type from Refs.~\cite{maxwell1,maxwell3}, thicker line type is an extension of that model including 
fits to $\sigma_{LT'}$ data from Ref.~\cite{sltp}). Note that this model is only available for the $K^+\Lambda$
final state and calculations go to a maximum $W$ of 2.275~GeV.
\item The Regge model of Guidal {\it et al.} (GLV)~\cite{glv} (green/dotted).
\item The Regge plus resonance model of Ghent (RPR)~\cite{corthals} (black/solid curves - RPR-2007 thinner 
line type, RPR-2011 thicker line type). For the $K^+\Sigma^0$ comparison, only the RPR-2007 version 
is presently available. 
\end{itemize}
  
A number of observations can be made independent of the model calculations: 

\begin{enumerate}
\item The production dynamics for $K^+\Lambda$ and $K^+\Sigma^0$ are quite different for
$W \le 2$~GeV. However, as $W$ increases further, the production mechanisms become similar. This
is to be expected as $KY$ production is known to be dominated by $t$-channel exchanges at higher energies.

\item The $K^+\Lambda$ production dynamics are dominated by $t$-channel exchange over the full
resonance region as indicated by the strong forward peaking of $\sigma_U$ in Figs.~\ref{lam_q1_ca}
and \ref{lam_q1_cb}. However, given the mid-angle peaking of $\sigma_U$ for $K^+\Sigma^0$ below 
2~GeV, clearly $s$-channel contributions play a much more significant role for this final state.

\item The forward peaking of $\sigma_U$ and $\sigma_{LT}$ for $K^+\Lambda$ compared to $K^+\Sigma^0$ 
can be qualitatively explained by the effect of the longitudinal coupling of the virtual photons. We 
note that the two channels are of nearly equal strength at $Q^2$=0~GeV$^2$~\cite{mccracken,dey}, 
while here at $Q^2$=1.80~GeV$^2$, the $K^+\Lambda$ channel is stronger than the $K^+\Sigma^0$ channel 
at forward angles by a factor of 3 to 4. For transverse (real) photons, the $t$-channel mechanism at 
low $t$ is dominated by vector $K^{*+}$ exchange, which relates directly to the magnitudes 
of the coupling constants $g_{K^*YN}$ relative to $g_{KYN}$. As $Q^2$ rises from zero, the photon 
can acquire a longitudinal polarization and the importance of pseudoscalar $K^+$ exchange increases. Given 
that $g_{K\Lambda N}^2 \gg g_{K\Sigma^0 N}^2$~\cite{adel_saghai,deswart}, this effect increases the cross 
section for $K^+\Lambda$ relative to $K^+\Sigma^0$ (this is consistent with the arguments presented in
Ref.~\cite{5st}). This argument is consistent with our observation of a sizable $\sigma_{LT}$ for 
$K^+\Lambda$ and a $\sigma_{LT}$ consistent with zero for $K^+\Sigma^0$. It should also be the case that 
since $g_{K^*\Sigma N} \gg g_{K\Sigma N}$, $K^*$ exchange should dominate the $K^+\Sigma^0$ channel. Because 
$K^*$ exchange must vanish at forward angles due to angular momentum conservation, the $K^+\Sigma^0$ cross 
section should also decrease at forward angles~\cite{glv}.

\item For $K^+\Lambda$, $\sigma_{TT}$ is consistent with zero up to about $W$=1.9~GeV then develops a
strong forward peaking that abruptly changes sign at about $W$=2.2~GeV. For $K^+\Sigma^0$, $\sigma_{TT}$
peaks at mid-range angles up to $W$=2~GeV and then looks very similar to $K^+\Lambda$ for higher $W$.
This higher $W$ response is well explained by the interference of the $K$ and $K^*$ Regge trajectories. 

\item For $K^+\Lambda$, $\sigma_{LT'}$ is relatively flat over the full angular range up to $W$=2~GeV and 
then develops a strong forward peaking for higher $W$ very similar to the other interference structure 
functions. We also note that it is significantly reduced at this $Q^2$ compared to the results at
$Q^2$=0.65 and 1.0~GeV$^2$ shown in Ref.~\cite{sltp}. $\sigma_{LT'}$ for $K^+\Sigma^0$ is consistent with 
zero over the full angular range.
\end{enumerate}

Comparing the data in Figs.~\ref{lam_q1_ca} to \ref{sig_q1_cb} to the different single-channel model
calculations, it is apparent that none of the models is successful at fully describing all of the data. 
A few general remarks are in order:

\begin{enumerate}
\item In general the models agree better with the $K^+\Lambda$ data than with the $K^+\Sigma^0$ data.
This likely arises, in part, due to the fact that better quality data for $K^+\Lambda$ is available than for
$K^+\Sigma^0$. However, as the resonance content is stronger in $K^+\Sigma^0$ compared to $K^+\Lambda$ for
$W < 2$~GeV given that the Regge predictions for $K^+\Lambda$ are in much closer agreement with the
$\sigma_U$ measurements compared to $K^+\Sigma^0$, the reaction mechanism for $K^+\Sigma^0$ is most certainly 
more complicated compared to $K^+\Lambda$, and thus more difficult to model correctly.

\item The models reproduce reasonably well the forward peaking strength in $\sigma_U$, $\sigma_{LT}$,
and $\sigma_{TT}$ for $K^+\Lambda$ and $K^+\Sigma^0$ for both final states for higher $W$. At
$W < 2$~GeV where the resonance contributions are a larger contribution relative to the non-resonant
background, the agreement is noticeably worse.

\item None of the models reproduces the trends in $\sigma_{LT'}$ for either final state across the
full $W$ spectrum. Interestingly, the hadrodynamic model of Maxwell {\it et al.} that includes the available
$\sigma_{LT'}$ data from Ref.~\cite{sltp} has by far the worst agreement with these data, although the
available $\sigma_{LT'}$ data only go up to $Q^2$=1.0~GeV$^2$.

\item The GLV Regge model that includes no $s$-channel resonance terms, does as well as any of the
other models in describing these data. For the $K^+\Sigma^0$ final state for $W < 2$~GeV, which has
strong $s$-channel contributions, the GLV model significantly underpredicts $\sigma_U$. However, for
$K^+\Lambda$, which has a much more significant $t$-channel exchange component within the resonance
region, the GLV model underpredicts $\sigma_U$ for $W < 1.9$~GeV. But for $W > 2.2$~GeV, 
the GLV model well matches the data for both final states over our full kinematic phase space.

\item For $K^+\Lambda$, the RPR-2011 model fares noticeably worse than for the RPR-2007 model over all
angles for $W < 2.1$~GeV for all of the structure functions. For higher $W$, where the response is
essentially fully $t$-channel, the RPR-2007 and RPR-2011 models agree well with the data and with
each other.
\end{enumerate}

\subsection{Energy Dependence}

\begin{sidewaysfigure*}[htbp]
\vspace{14.0cm}
\includegraphics{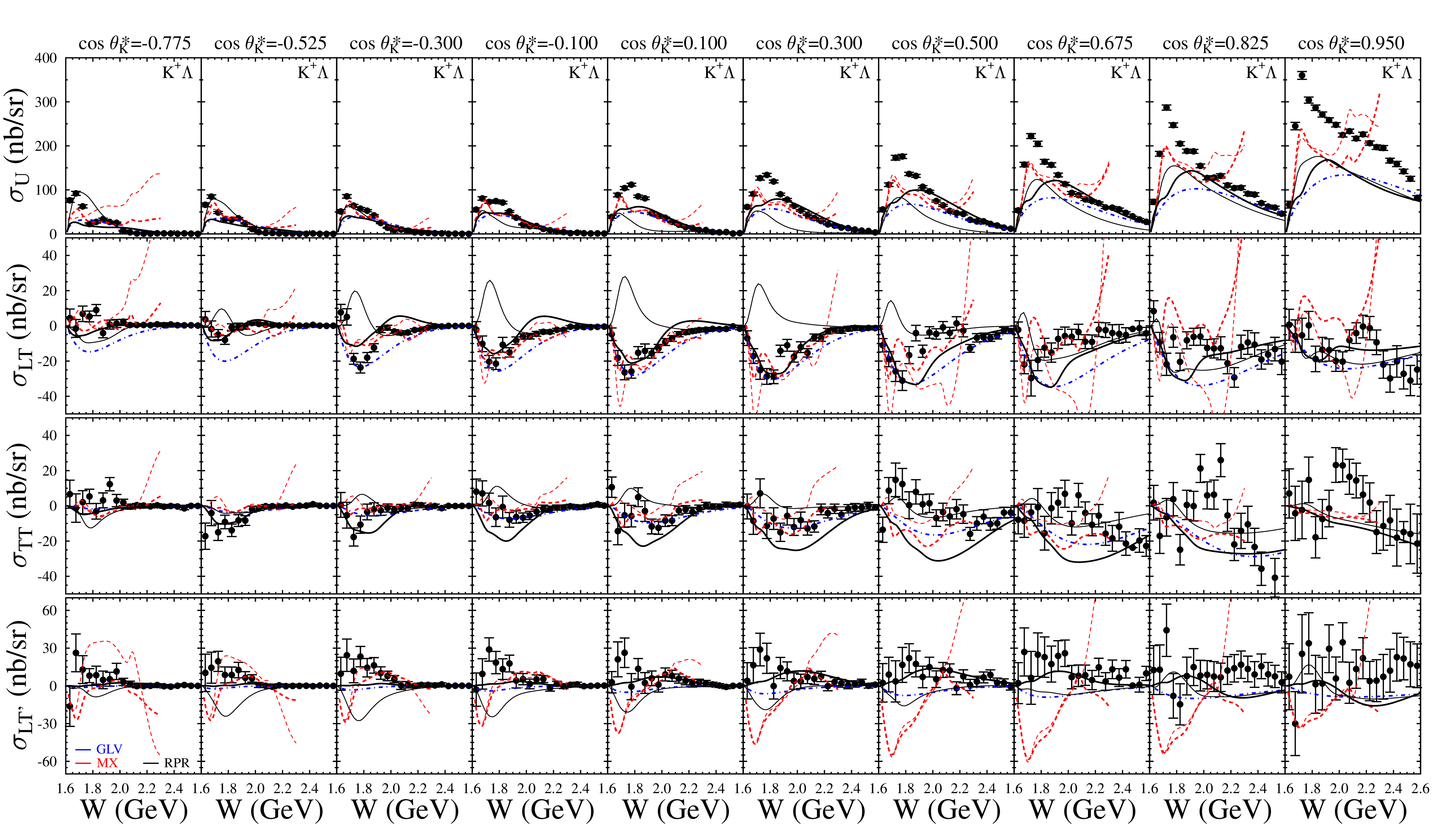}
\caption{(Color online) Structure functions $\sigma_U$, $\sigma_{LT}$, $\sigma_{TT}$, and $\sigma_{LT'}$ 
(in nb/sr) for $K^+\Lambda$ production vs. $W$ at 5.499~GeV for $Q^2$=1.80~GeV$^2$ and for the 10 
$\cos \theta_K^*$ values. The error bars represent the statistical uncertainties only. The curves are 
defined in the caption of Fig.~\ref{lam_q1_ca}.}
\label{lam_q1_w} 
\end{sidewaysfigure*}

\begin{sidewaysfigure*}[htbp]
\vspace{14.0cm}
\includegraphics{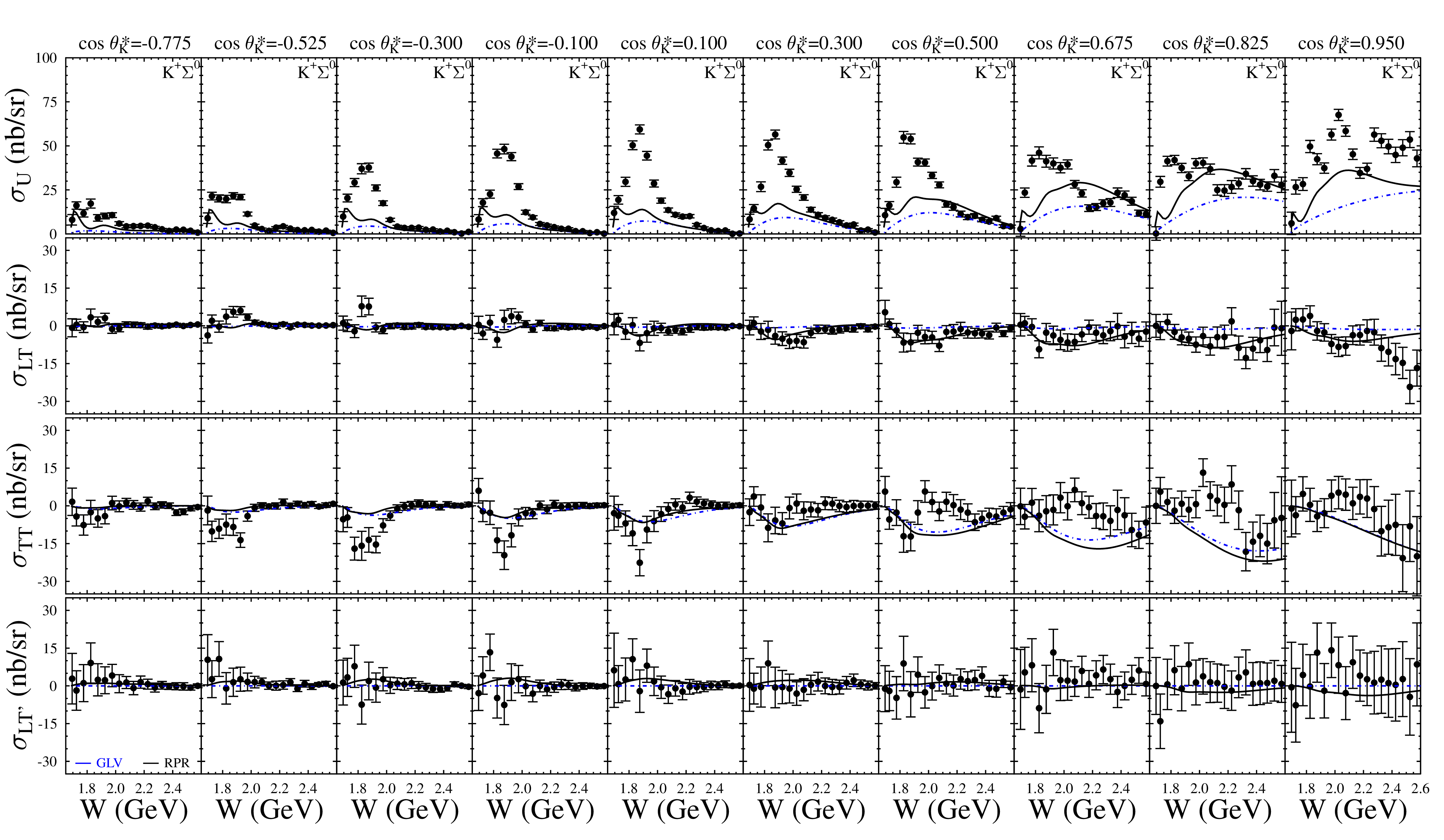}
\caption{(Color online) Structure functions $\sigma_U$, $\sigma_{LT}$, $\sigma_{TT}$, and $\sigma_{LT'}$ 
(in nb/sr) for $K^+\Sigma^0$ production vs. $W$ at 5.499~GeV for $Q^2$=1.80~GeV$^2$ and for the 10 
$\cos \theta_K^*$ values. The error bars represent the statistical uncertainties only. The curves are 
defined in the caption of Fig.~\ref{lam_q1_ca}.}
\label{sig_q1_w} 
\end{sidewaysfigure*}

To more directly look for $s$-channel resonance evidence, the extracted structure functions are 
presented as a function of the center-of-mass energy $W$ for our ten values of $\cos \theta_K^*$. 
Figs.~\ref{lam_q1_w} and \ref{sig_q1_w} show the results for our $K^+\Lambda$ and $K^+\Sigma^0$ data, 
respectively, at $Q^2$=1.80~GeV$^2$. 

A number of observations can be made regarding the data:

\begin{enumerate}
\item For $K^+\Lambda$ production, $\sigma_U$ shows a broad peak at about 1.7~GeV at forward angles, 
and two peaks separated by a dip at about 1.75~GeV for our two backward angle points. This corroborates 
similar features seen in recent photo- and electroproduction results~\cite{saphir1,saphir2,mccracken,5st,bradford-cs}. 
Within existing hadrodynamic models, the structure just above the threshold region is typically accounted for 
by the known $N(1650)1/2^-$, $N(1710)1/2^+$, and $N(1720)3/2^+$ nucleon resonances. However, there is no 
consensus as to the origin of the bump feature at $\sim$1.9~GeV that was first seen in the $K^+\Lambda$ 
photoproduction data from SAPHIR~\cite{saphir1}. It is tempting to speculate that this is evidence for a 
previously ``missing'', negative-parity $J=3/2$ resonance at 1.96~GeV predicted in the quark model of Capstick 
and Roberts~\cite{capstick}. This explanation was put forward in the work of Bennhold and Mart~\cite{bennhold}, 
in which they postulated the existence of a $3/2^-$ state at 1.9~GeV. However, in Ref.~\cite{nikonov} it was
shown that a $N(1900)3/2^+$ state is required to explain the beam-recoil polarization data for $K^+\Lambda$. 
In Ref.~\cite{saghai} this broad bump in the $K^+\Lambda$ cross section could be explained by accounting for 
$u$-channel hyperon exchanges.

\item For $K^+\Lambda$, $\sigma_{LT}$ has about 20\% of the strength of $\sigma_U$ and is consistently
negative. For $K^+\Sigma^0$, $\sigma_{LT}$ is nearly zero everywhere except for $W$=1.9~GeV at back angles.

\item The $\sigma_{TT}$ structure function is quite similar for $K^+\Lambda$ and $K^+\Sigma^0$ over all
kinematics with a strength comparable to $\sigma_{LT}$.

\item For $K^+\Lambda$, $\sigma_{LT'}$ shows significant structure for $W$ below 2.2~GeV. For higher $W$
it is consistent with zero.

\item In the $K^+\Sigma^0$ channel, $\sigma_U$ is peaked at about 1.9~GeV, which also matches the photoproduction 
result~\cite{saphir2,dey,bradford-cs}. $\sigma_{TT}$, while small, shows a broad feature in this same region. These 
features are consistent with a predominantly $s$-channel production mechanism. In this region, beyond the 
specific $N^*$ resonances believed to contribute to $K^+\Lambda$ production (and hence are strong candidates 
to contribute to $K^+\Sigma^0$ production), there are a number of known $\Delta^*$ resonances near 1.9~GeV
\cite{pdg} that can contribute to the $K^+\Sigma^0$ final state, particularly the $\Delta(1900)1/2^-$ and 
$\Delta(1910)1/2^+$. These $\Delta^*$ states are forbidden to couple to the $K^+\Lambda$ state due to isospin 
conservation.
\end{enumerate}

The comparisons of the model calculations to the data clearly indicate that significant new constraints on 
the model parameters will be brought about when these new electroproduction data are included in the fits. 
We conclude that the $W$ dependence of $K^+\Lambda$ and $K^+\Sigma^0$ production provides strong evidence 
for baryon resonance activity within the reaction mechanism, but that the data in comparison to present models 
do not allow any simple statement to be made. We further conclude that at the current time the models that are
limited to fits of the photoproduction data only, cannot adequately describe the electroproduction data.

\subsection{$Q^2$ Dependence}

\begin{figure*}[htbp]
\vspace{8.2cm}
\includegraphics{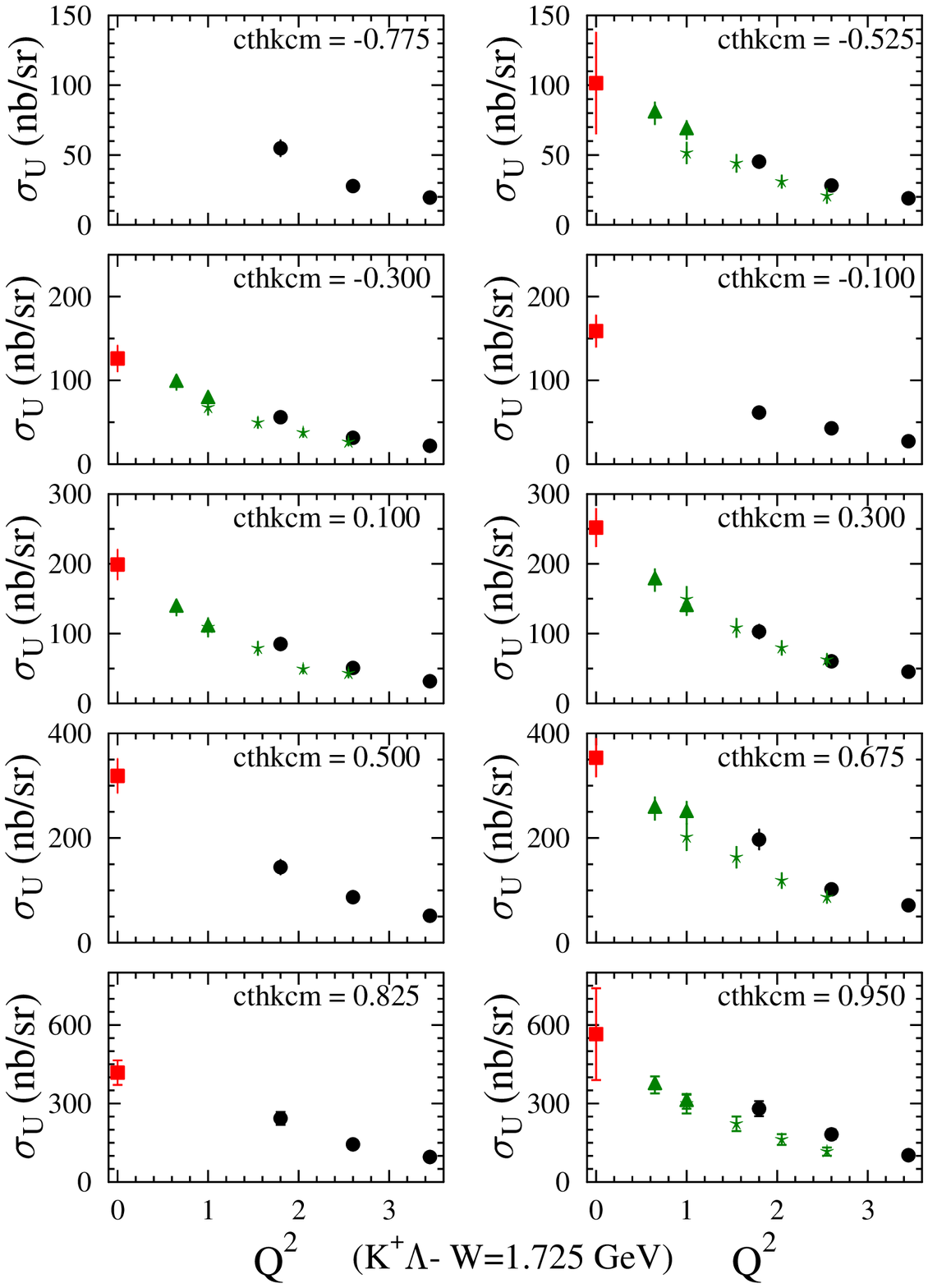}
\includegraphics{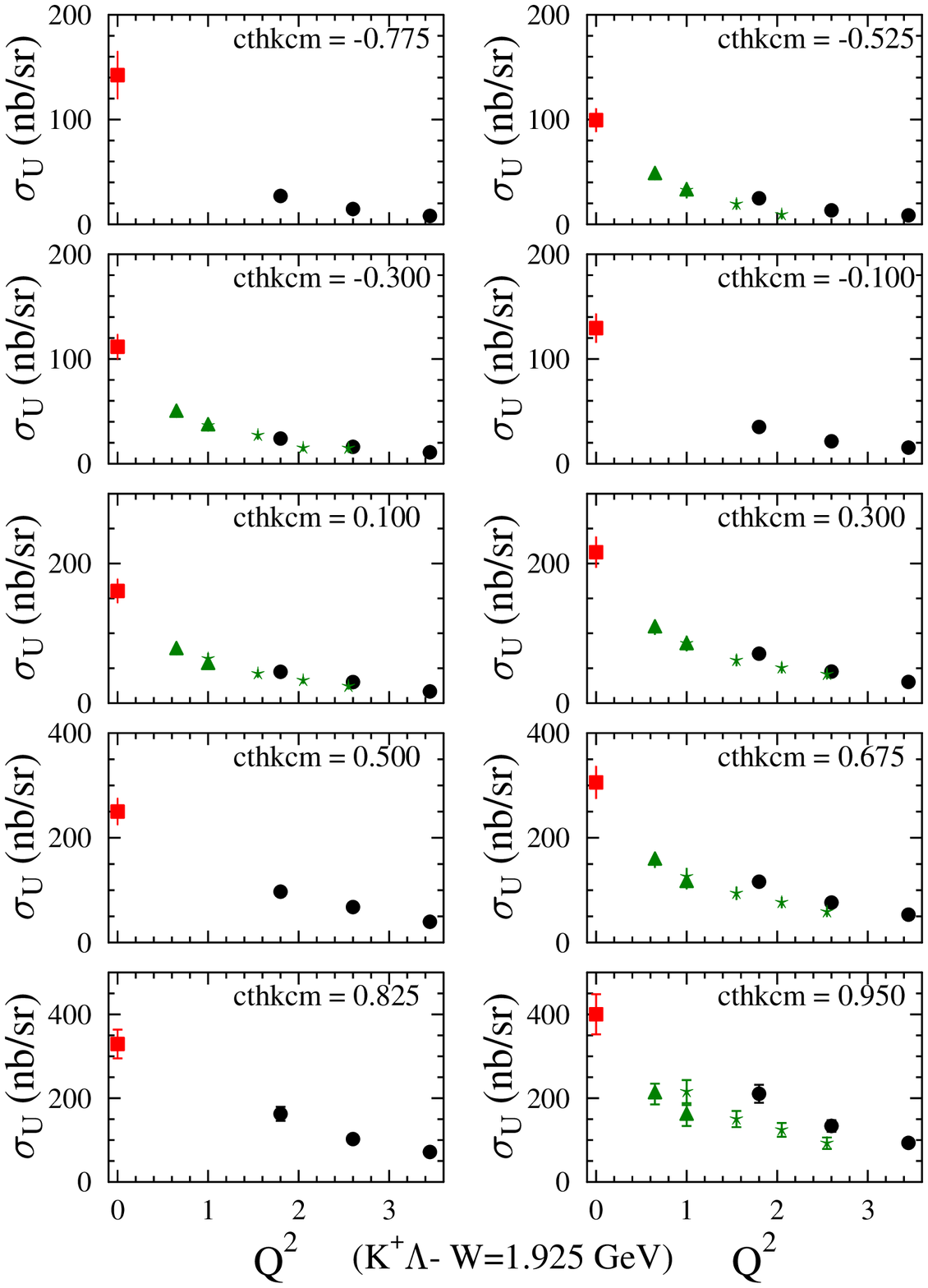}
\caption{(Color online) Structure function $\sigma_U$ vs. $Q^2$ (GeV$^2$) for the $K^+\Lambda$ final state 
for two values of $W$=1.725 and 1.925~GeV. The black circles are the data from this work, the red squares 
are the photoproduction points from Ref.~\cite{mccracken}, and the green stars and triangles are from the
lower $Q^2$ data from Ref.~\cite{5st}. The error bars include both statistical and systematic uncertainties.}
\label{lam-q2} 
\end{figure*}

\begin{figure*}[htbp]
\vspace{8.2cm}
\includegraphics{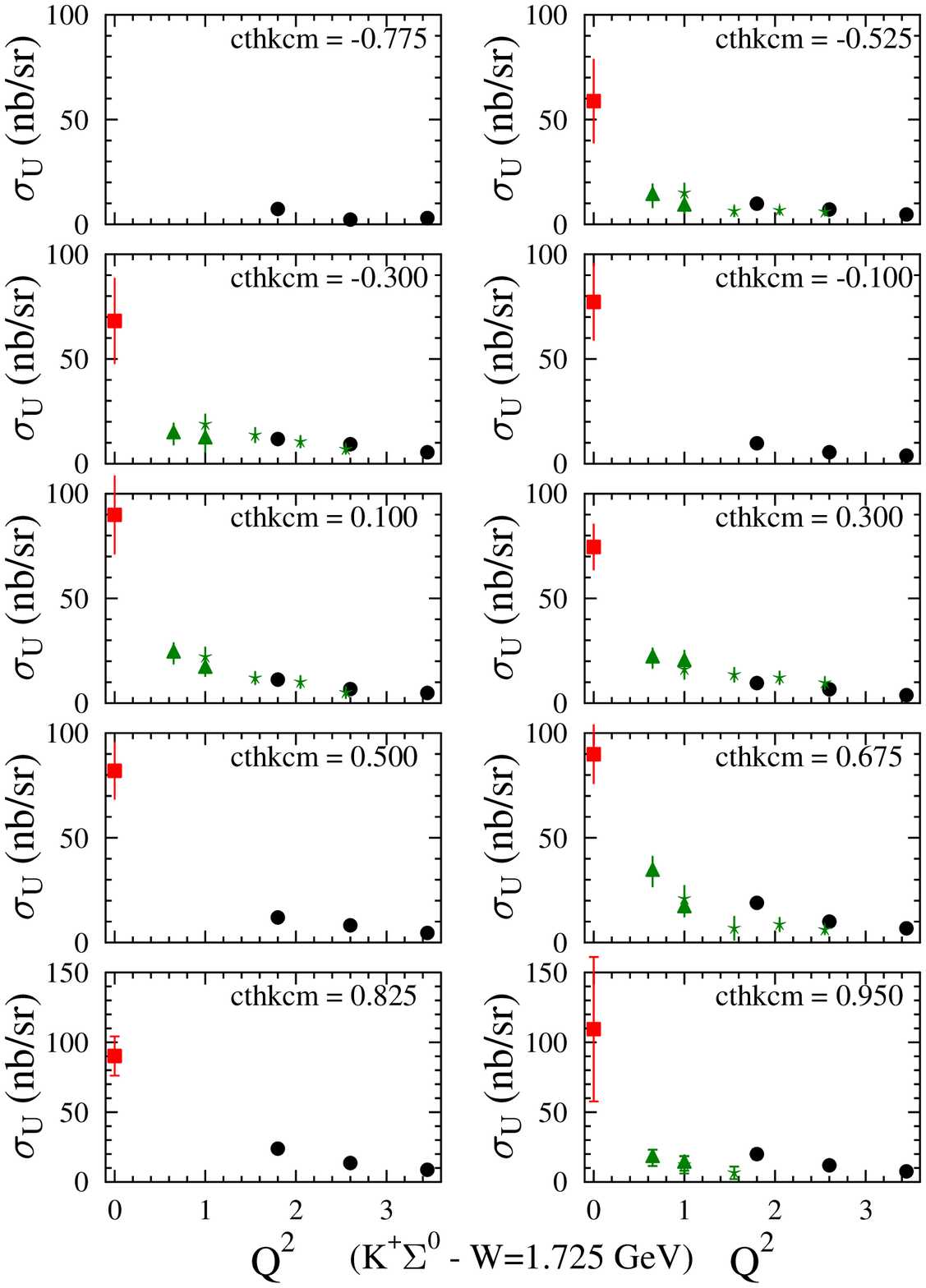}
\includegraphics{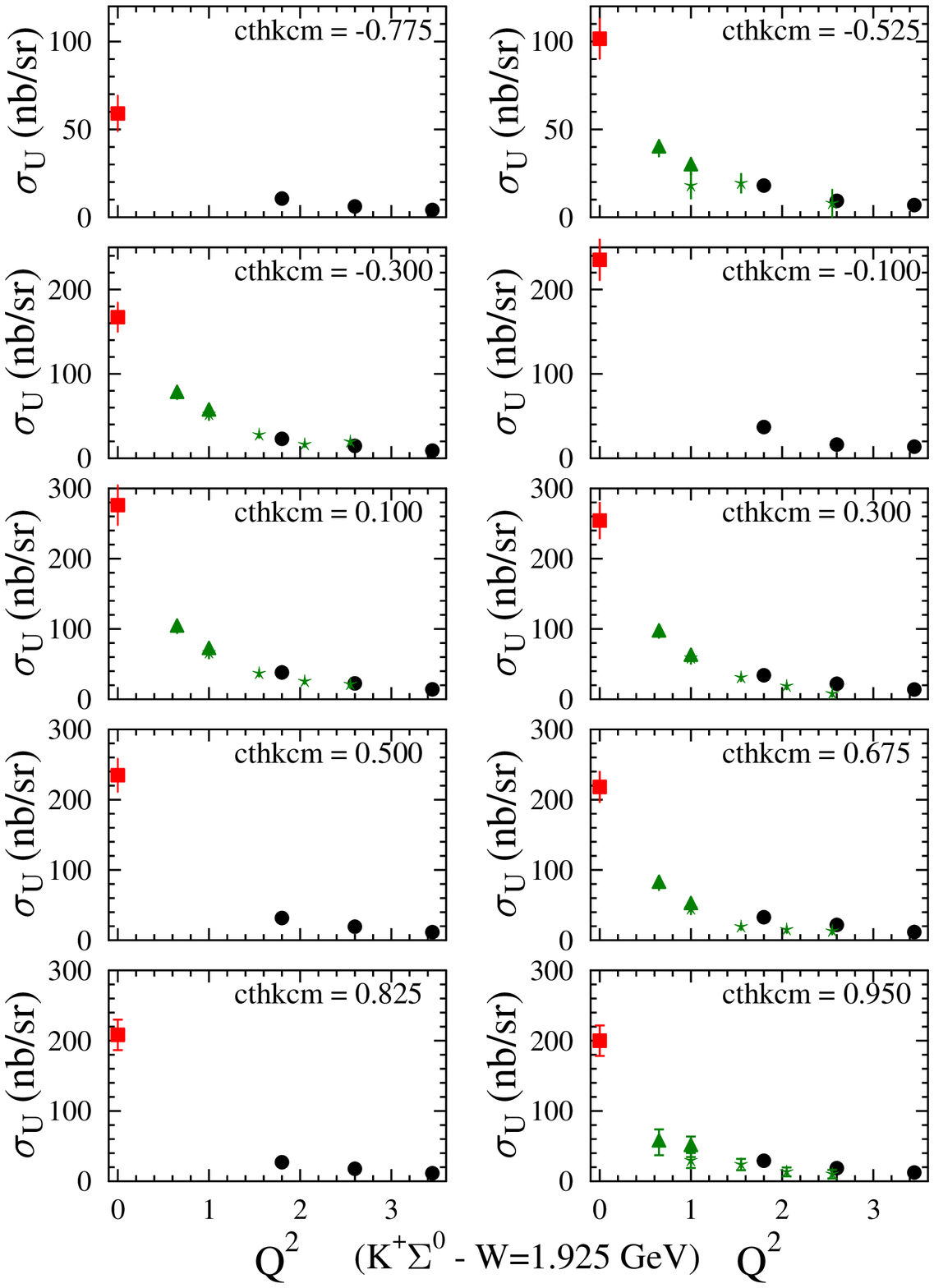}
\caption{(Color online) Structure function $\sigma_U$ vs. $Q^2$ (GeV$^2$) for the $K^+\Sigma^0$ final state 
for two values of $W$=1.725 and 1.925~GeV. The black circles are the data from this work, the red squares 
are the photoproduction points from Ref.~\cite{dey}, and the green stars and triangles are from the
lower $Q^2$ data from Ref.~\cite{5st}. The error bars include both statistical and systematic uncertainties.}
\label{sig-q2} 
\end{figure*}

Our data set provides a large $Q^2$ reach and it is instructive to study the $W$ spectra for increasing values 
of $Q^2$.  These data are shown in Figs.~\ref{lam-q2} and \ref{sig-q2} for the $K^+\Lambda$ and $K^+\Sigma^0$
final states at two representative $W$ points, 1.725 and 1.925~GeV. Included on these plots are the
photoproduction differential cross sections for $K^+\Lambda$ from Ref.~\cite{mccracken} and $K^+\Sigma^0$
from Ref.~\cite{dey} at $Q^2$=0 for the kinematic points where they are available. Also shown are the data
from $\sigma_U$ from Ref.~\cite{5st} from two different data sets, (i). $E_b$=2.567~GeV, $Q^2$=0.65, 1.0~GeV$^2$
and (ii). $E_b$=4.056~GeV, $Q^2$=1.0, 1.55, 2.05, 2.55~GeV$^2$ at kinematic points that are reasonably close
to the present data.

What is seen by studying the $Q^2$ evolution of $\sigma_U$ is a reasonably smooth fall-off from the
photon point. As the photoproduction data involve a purely transverse response, this smooth fall-off to
finite $Q^2$ in these kinematics predominantly indicates a small longitudinal response. This is also 
indicated by the small strengths of $\sigma_{LT}$ and $\sigma_{LT'}$ relative to $\sigma_U$ in 
Figs.~\ref{lam_q1_ca} to \ref{sig_q1_w} for back- and mid-range angles for the $K^+\Lambda$ final state
and for all angles for the $K^+\Sigma^0$ final state. However, there is clearly a non-negligible longitudinal
response in the $K^+\Lambda$ data at forward angles and for higher $W$ as seen in these data (and also
seen in the data of Ref.~\cite{5st}). Note that the comparisons shown in Figs.~\ref{lam-q2} and \ref{sig-q2} 
are only for qualitative comparisons as the kinematics are not a perfect match in all cases from 
Refs.~\cite{mccracken,dey,5st} to the present data.

The smooth fall-off of $\sigma_U$ with increasing $Q^2$ is consistent with the findings of the lower
$Q^2$ analysis of $K^+\Lambda$ and $K^+\Sigma^0$ electroproduction from Ref.~\cite{5st}. As was the case
in that work, it is seen that the interference structure functions $\sigma_{LT}$, $\sigma_{TT}$, and
$\sigma_{LT'}$ for both final states do not demonstrate any strong $Q^2$ dependence. However, detailed
comparisons with available models will be important to gain insight into the associated form factors
for the $N^*$ resonances found from fits to the photoproduction data.

\subsection{Legendre Fits}

In order to investigate the possible evidence for the presence of $s$-channel resonance contributions in
the separated structure functions, we have considered two different approaches. The first is with a fit 
of the individual structure functions $\sigma_U$, $\sigma_{LT}$, $\sigma_{TT}$, and $\sigma_{LT'}$ versus 
$\cos \theta_K^*$ for each $Q^2$ and $W$ point for the $K^+\Lambda$ and $K^+\Sigma^0$ final states using 
a truncated series of Legendre polynomials as:
\begin{equation}
\label{incoherent}
C_{\ell = 0 \to 3} = \int_{-1}^{+1} \frac{d\sigma_{U,LT,TT,LT'}}{d\Omega^*} P_\ell(\cos \theta_K^*)~d\!\cos \theta_K^*.
\end{equation}

\begin{figure}[htbp]
\vspace{7.5cm}
\includegraphics{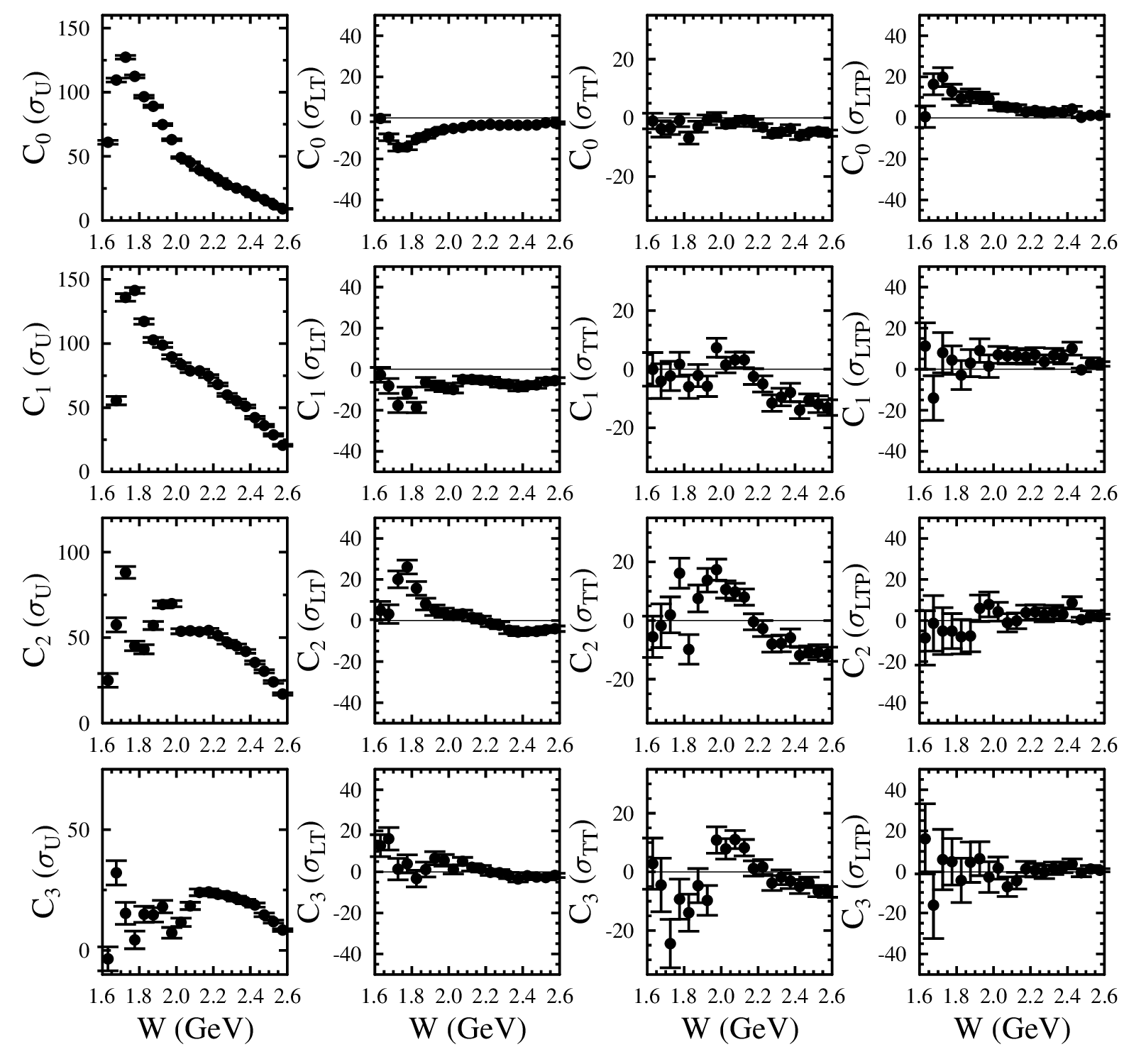}
\caption{Legendre polynomial fit coefficients (nb) from Eq.(\ref{incoherent}) vs. $W$ for the $K^+\Lambda$ 
separated structure functions $\sigma_U$, $\sigma_{LT}$, $\sigma_{TT}$, and $\sigma_{LT'}$ for 
$Q^2$=1.80~GeV$^2$.}
\label{lam-leg-incoherent} 
\end{figure}

\begin{figure}[htbp]
\vspace{7.5cm}
\includegraphics{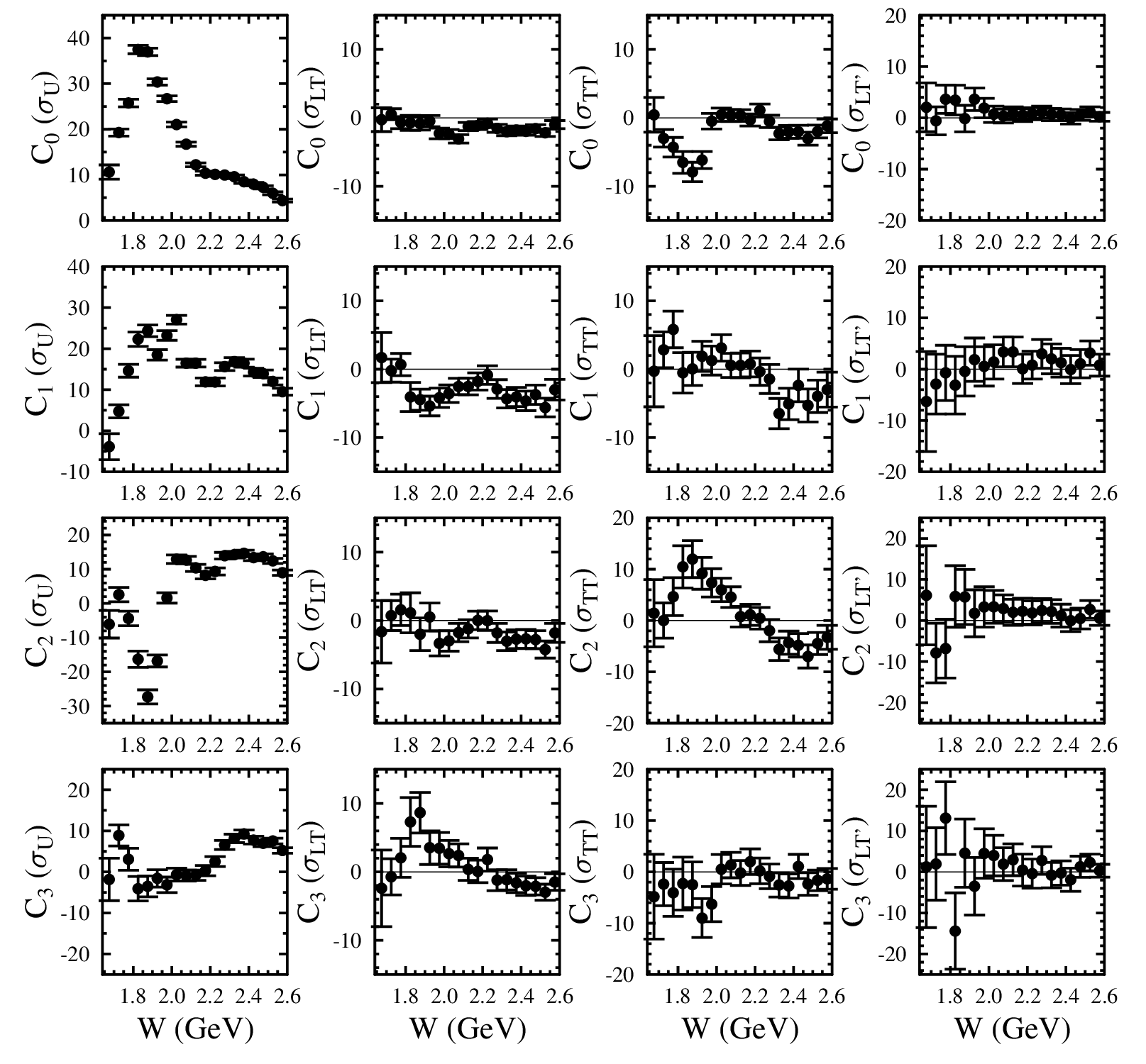}
\caption{Legendre polynomial fit coefficients (nb) from Eq.(\ref{incoherent}) vs. $W$ for the $K^+\Sigma^0$ 
separated structure functions $\sigma_U$, $\sigma_{LT}$, $\sigma_{TT}$, and $\sigma_{LT'}$ for 
$Q^2$=1.80~GeV$^2$.}
\label{sig-leg-incoherent} 
\end{figure}

The fit coefficients for $\ell = 0 \to 3$ are shown for $K^+\Lambda$ in Fig.~\ref{lam-leg-incoherent} and
for $K^+\Sigma^0$ in Fig.~\ref{sig-leg-incoherent} for $Q^2=1.80$~GeV$^2$. The structures seen in these
coefficients versus $W$ are likely indicative of $s$-channel contributions. Note that the appearance of a
structure at a given value of $W$ in each of the different $C_\ell$ coefficients most likely suggests the 
presence of a dynamical effect rather than the signature of an $N^*$ contribution. Instead, the appearance 
of a structure in a single $C_\ell$ coefficient at the same $W$ value and in each of the $Q^2$ points is 
more likely a signal of an $N^*$ contribution.

The fits for $K^+\Lambda$ show structures at $W$=1.7~GeV in $C_0$ for both $\sigma_U$ and $\sigma_{LT}$, 
$W$=1.9~GeV in $C_2$ and $C_3$ for $\sigma_U$, and $W$=2.2~GeV in $C_3$ for $\sigma_U$. The fits for 
$K^+\Sigma^0$ show structures at $W$=1.9~GeV in $C_0$ and $C_2$ for $\sigma_U$ and $\sigma_{TT}$. Of course, 
making statements regarding the possible orbital angular momentum of the associated $s$-channel 
resonances requires care as interference effects among the different partial waves can cause strength for a 
given orbital angular momentum value to be spread over multiple Legendre coefficients.

\begin{figure}[htbp]
\vspace{7.3cm}
\includegraphics{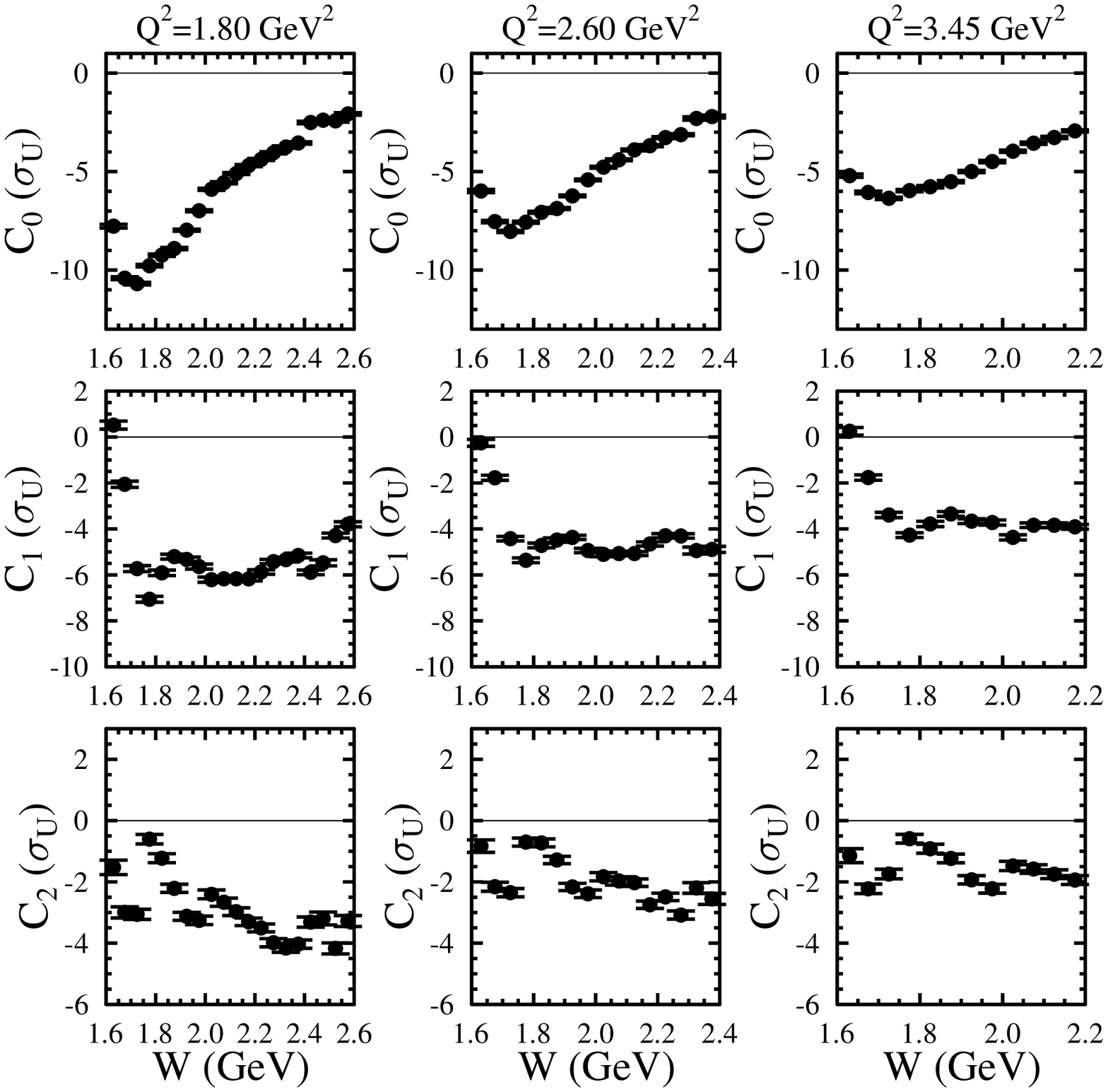}
\caption{Coherent Legendre polynomial fit coefficients ((nb/sr)$^{1/2}$) from Eq.(\ref{coherent}) vs. $W$ 
for the $K^+\Lambda$ separated structure function $\sigma_U$ for $Q^2$=1.80, 2.60, and 3.45~GeV$^2$.}
\label{lam-leg-coherent} 
\end{figure}

\begin{figure}[htbp]
\vspace{7.3cm}
\includegraphics{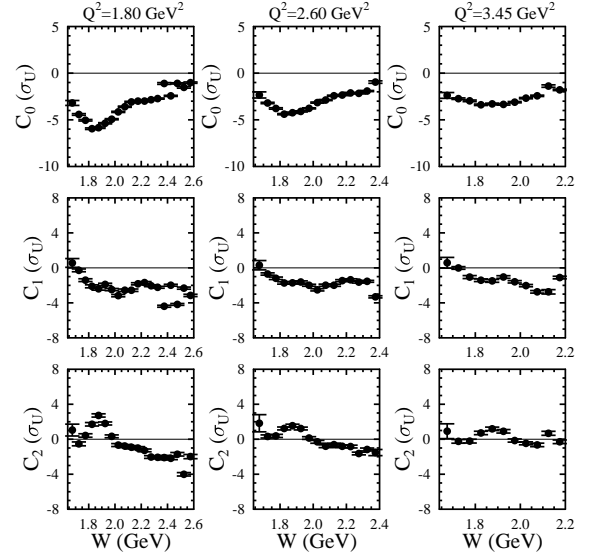}
\caption{Coherent Legendre polynomial fit coefficients ((nb/sr)$^{1/2}$) from Eq.(\ref{coherent}) vs. $W$ 
for the $K^+\Sigma^0$ separated structure function $\sigma_U$ for $Q^2$=1.80, 2.60, and 3.45~GeV$^2$.}
\label{sig-leg-coherent} 
\end{figure}

In a second approach, each of the Legendre coefficients can be further expanded in terms of products of pairs 
of multipole amplitudes, but these expansions quickly become unwieldy as the number of participating partial 
waves increases. However, one simple thing that can be done for additional insight is to fit the structure 
functions with a coherent Legendre series of the form:
\begin{equation}
\label{coherent}
\frac{d\sigma_{U,LT,TT,LT'}}{d\Omega^*} = \left[\sum_{\ell=0}^{2} C_\ell(Q^2,W) P_\ell(\cos \theta_K^*)\right]^2 + C_x^2.
\end{equation}
\noindent
Here the $P_\ell$ are the usual Legendre polynomials. The coefficients $C_\ell(Q^2,W)$ are the amplitudes
of the coherent $S$, $P$, and $D$-wave contributions, respectively, while $C_x$ takes into account a
incoherent ``background'' connected with higher-order terms that are not taken into account in the
truncated sum. Of course, one must take care against making too much of the fit results using the
simplistic form of Eq.(\ref{coherent}). This approach is not meant to be an attempt at a true amplitude fit.
Rather the point is to look for structures that appear at a given $W$ and for each $Q^2$ for a given $C_\ell$ 
coefficient as suggestive evidence for possible $N^*$ resonance contributions. Fig.~\ref{lam-leg-coherent} 
shows the Legendre coefficient from this approach for $\sigma_U$ for the $K^+\Lambda$ reaction for the three 
$Q^2$ points in this analysis. Fig.~\ref{sig-leg-coherent} is the corresponding figure for $K^+\Sigma^0$.

The fit coefficients for $\sigma_U$ shown in Figs.~\ref{lam-leg-coherent} and \ref{sig-leg-coherent}
show reasonable correspondence among all three $Q^2$ points. For the $K^+\Lambda$ fits, strength is
seen at: $W$=1.7~GeV in $C_0$, $W$=1.9~GeV in $C_1$, and $W$=2.2~GeV in $C_2$. While it might be tempting
to view this as corroboration of the findings of the $K^+\Lambda$ photoproduction amplitude analysis from
Ref.~\cite{sch-sarg}, obviously more detailed work is required. For the $K^+\Sigma^0$ fits, strength is
seen at $W$=1.85~GeV in $C_0$ and $W$=1.9~GeV in $C_2$. It is interesting that there is no signature of
strength in the $P$-wave as seen through the coefficient $C_1$, but again a higher-order analysis will be 
required to make more definite statements.

\section{Summary and Conclusions}
\label{conclusions}

We have measured $K^+\Lambda$ and $K^+\Sigma^0$ electroproduction off the proton over a wide range of kinematics 
in the nucleon resonance region. We have presented data for the differential cross sections and separated 
structure functions $\sigma_U$, $\sigma_{LT}$, $\sigma_{TT}$, and $\sigma_{LT'}$ for $Q^2$ from 1.4
to 3.9~GeV$^2$, $W$ from threshold to 2.6~GeV, and spanning nearly the full center-of-mass angular range for
the $K^+$. In addition to the increased kinematic reach of these data relative to the previously 
published $K^+Y$ electroproduction structure functions from CLAS in Ref.~\cite{5st}, this new data set is
an order of magnitude larger, allowing for finer binning in $W$ and $\cos \theta_K^*$.

The structure function data for both $K^+\Lambda$ and $K^+\Sigma^0$ indicates that for $W$ below 2.2~GeV
and back angles, there is considerable strength of contributing $s$-channel resonances for $K^+\Lambda$ and
$K^+\Sigma^0$. For higher $W$, the $t$-channel non-resonant background dominates and the reaction dynamics 
are well described solely through interference of $K$ and $K^*$ Regge trajectories.

A Legendre analysis confirms these qualitative statements. For the $K^+\Lambda$ final state,
the Legendre moments of the structure functions indicate possible $s$-channel resonant contributions
in the $S$-wave near 1.7~GeV, in the $P$-wave near 1.9~GeV, and in the $D$-wave near 2.2~GeV. This
is in qualitative agreement with the more detailed amplitude analysis of Ref.~\cite{sch-sarg}. For
the $K^+\Sigma^0$ final state, strong $S$-wave strength is seen at 1.8~GeV and strong $D$-wave strength
is seen above 1.9~GeV, precisely where several $\Delta^*$ states are expected to couple. Of course
more detailed and quantitative statements await including these data into the coupled-channel partial
wave fits. Such analyses would help to provide important complementary cross checks to the fit results of
the recent Bonn-Gatchina coupled-channels results from Ref.~\cite{bonngat1} that seem to favor a much
richer mix of states to describe the available photoproduction data.

Finally, detailed comparisons of our data have been made with several existing models. These include
the hadrodynamic model of Maxwell {\it et al.}~\cite{maxwell3} that has been constrained by both the 
CLAS photo- and electroproduction data sets (both cross sections and spin observables), the Regge model 
of Guidal {\it et al.}~\cite{glv} that has only been constrained by high-energy photoproduction data to 
fix the parameters of the Regge trajectories, and the Regge plus resonance model from Ghent~\cite{corthals} 
that has been constrained by the existing high statistics photoproduction data.
None of the available models does a satisfactory job of describing the structure functions below
$W = 2$~GeV for either $K^+\Lambda$ or $K^+\Sigma^0$. In fact, several of the more recent models
(e.g. RPR-2011 and the MX model including the CLAS $\sigma_{LT'}$ data) actually are in worse agreement
with the data below 2~GeV than for earlier versions of the models. Clearly more work on the modeling
and possibly the fitting/convergence algorithms is required to be able to fully understand the contributing 
$N^* \to K^+\Lambda$ and $N^*, \Delta^* \to K^+\Sigma^0$ states and to reconcile the results from the 
single-channels models with the currently available coupled-channel models.

\section*{Acknowledgments}

We would like to acknowledge the outstanding efforts of the staff of the Accelerator and the Physics Divisions 
at JLab that made this experiment possible. This work was supported in part by the U.S. Department of Energy,
the National Science Foundation, the Italian Istituto Nazionale di Fisica Nucleare, the French Centre National 
de la Recherche Scientifique, the French Commissariat \`{a} l'Energie Atomique, the United Kingdom's Science
and Technology Facilities Council, the Chilean Comisi\'{o}n Nacional de Investigaci\'{o}n Cient\'{i}fica y 
Tecnol\'{o}gica (CONICYT), and the National Research Foundation of Korea. The Southeastern Universities Research 
Association (SURA) operated Jefferson Lab under United States DOE contract DE-AC05-84ER40150 during this work.


\begin{thebibliography}{99}

\bibitem{burkert}
V.D. Burkert and T.-S.H. Lee, Int. J. Phys. E {\bf 13}, 1035 (2004).


\bibitem{dudek}
J.J. Dudek and R. Edwards, Phys. Rev. D {\bf 85}, 054016 (2012).

\bibitem{oset}
E. Oset {\it et al.}, Int. J. Mod. Phys. A {\bf 20}, 1619 (2005).

\bibitem{capstick}
S. Capstick and W. Roberts, Phys. Rev. D {\bf 58}, 074011 (1998).

\bibitem{isgur} 
N. Isgur, Proceedings of the NSTAR 2000 Conference, eds. V.D. Burkert, L. Elouadrhiri, J.J. Kelly, 
and R. Minehart, (World Scientific, Singapore, 2001), p. 403.

\bibitem{buluva}
J.M. Buluva {\it et al.}, Phys. Rev. D {\bf 79}, 034505 (2009).

\bibitem{bonngat1}
A.V. Anisovich {\it et al.}, Eur. Phys. J A {\bf 48}, 15 (2012).

\bibitem{bonngat2}
A.V. Anisovich {\it et al.}, arXiv:1205.2255, (2012).

\bibitem{penner}
G. Penner and U. Mosel, Phys. Rev. C {\bf 66}, 055212 (2002).

\bibitem{julich}
M. D\"{o}ring {\it et al.}, Nucl. Phys. A {\bf 851}, 58 (2011).

\bibitem{ebac1}
H. Kamano {\it et al.}, Phys. Rev. C {\bf 81}, 065208 (2010).

\bibitem{said}
R.A. Arndt {\it et al.}, Phys. Rev. C {\bf 53}, 430 (1996); Int. J. Mod. Phys. A {\bf 18}, 449 (2003).

\bibitem{pdg}
J. Beringer {\it et al.} (PDG), Phys. Rev. D {\bf 86}, 010001 (2012).

\bibitem{carman_1} 
D.S. Carman {\it et al.} {\it (CLAS Collaboration)}, Phys. Rev. Lett. {\bf 90}, 131804 (2003).

\bibitem{carman_2} 
D.S. Carman {\it et al.} {\it (CLAS Collaboration)}, Phys. Rev. C {\bf 79}, 065205 (2009).

\bibitem{5st}
P. Ambrozewicz {\it et al.} {\it (CLAS Collaboration)}, Phys. Rev. C {\bf 75}, 045203 (2007).

\bibitem{sltp}
R. Nasseripour {\it et al.} {\it (CLAS Collaboration)}, Phys. Rev. C {\bf 77}, 065208 (2008).

\bibitem{sarantsev}
A.V. Sarantsev {\it et al.}, Eur. Phys. J. A {\bf 25}, 441 (2005).

\bibitem{shklyar}
V. Shklyar, H. Lenske, and U. Mosel, Phys. Rev. C {\bf 72}, 015210 (2005).

\bibitem{mart1}
T. Mart, AIP Conf. Proc. {\bf 1056}, 31 (2008).

\bibitem{nikonov}
V.A. Nikonov {\it et al.}, Phys. Lett. B {\bf 662}, 245 (2008).

\bibitem{bradford}
R.K. Bradford {\it et al.} {\it (CLAS Collaboration)}, Phys. Rev. C {\bf 75}, 035205 (2007).

\bibitem{klempt}
E. Klempt and R. Workman, \url{https://pdg.web.cern.ch/pdg/2012/reviews/rpp2012-rev-n-delta-resonances.pdf}.

\bibitem{diaz}
B. Julia-Diaz {\it et al.}, Nucl. Phys. A {\bf 755}, 463 (2005); B. Julia-Diaz {\it et al.}, Phys. Rev. C 
{\bf 73}, 055204 (2006).

\bibitem{martsul}
T. Mart and A. Sulaksono, Phys.Rev. C {\bf 74}, 055203  (2006); T. Mart and A. Sulaksono, nucl-th/0701007, 
(2007).

\bibitem{corthals}
T. Corthals {\it et al.}, Phys. Lett. B {\bf 656}, 186 (2007).

\bibitem{maxwell1}
O. Maxwell, Phys. Rev. C {\bf 85}, 034611 (2012).

\bibitem{janssen}
S. Janssen {\it et al.}, Phys. Rev. C {\bf 67}, 052201 (R) (2003).

\bibitem{mart2}
T. Mart, Eur. Phys. J. Web Conf. {\bf 3}, 07002 (2010).

\bibitem{indpol}
M. Gabrielyan {\it et al.}, AIP Conf. Proc. {\bf 1432}, 375 (2011).

\bibitem{harry-lee}
T.-S.H. Lee, private communication, (2012).

\bibitem{thoma}
U. Thoma, private communication, (2012).

\bibitem{maxwell2}
O. Maxwell, Phys. Rev. C {\bf 76}, 014621 (2007).

\bibitem{maxwell3}
A. de la Puente, O. Maxwell, and B.A. Raue, Phys. Rev. C {\bf 80}, 065205 (2009).

\bibitem{glv}
M. Guidal, J.M. Laget, and M. Vanderhaegen, Nucl. Phys. A {\bf 627}, 645 (1997);  Phys. Rev. C {\bf 61}, 025204 (2000);
Phys. Rev. C {\bf 68}, 058201 (2003).

\bibitem{decruz}
L. De Cruz {\it et al.}, Phys. Rev. Lett. {\bf 108}, 182002 (2012); L. De Cruz {\it et al.}, arXiv:1205.2195 (2012).

\bibitem{akerlof} 
C.W. Akerlof {\it et al.}, Phys. Rev. {\bf 163}, 1482 (1967).

\bibitem{boffi} 
S. Boffi, C. Giusti, and F.D. Pacati, Phys. Rep. {\bf 226}, 1 (1993); S. Boffi, C. Giusti, and F.D. Pacati, 
Nucl. Phys. A {\bf 435}, 697 (1985).

\bibitem{mecking}
B.A. Mecking {\it et al.}, Nucl. Inst. and Meth. A {\bf 503}, 513 (2003).

\bibitem{dcnim}
M.D. Mestayer {\it et al.}, Nucl. Inst. and Meth. A {\bf 449}, 81 (2000).

\bibitem{ccnim}
G.S. Adams {\it et al.}, Nucl. Inst. and Meth. A {\bf 465}, 414 (2001).

\bibitem{scnim}
E.S. Smith {\it et al.}, Nucl. Inst. and Meth. A {\bf 432}, 265 (1999).

\bibitem{ecnim}
M. Amarian {\it et al.}, Nucl. Inst. and Meth. A {\bf 460}, 239 (2001).

\bibitem{fsgen}
S. Stepanyan, FSGEN phase space event generator, private communication, (2011).

\bibitem{genev}
R. De Vita, Genova event generator, private communication, (2012).

\bibitem{mccracken}
M.E. McCracken {\it et al.} {\it (CLAS Collaboration)}, Phys. Rev. C {\bf 81}, 025201 (2010).

\bibitem{dey}
B. Dey {\it et al.} {\it (CLAS Collaboration)}, Phys. Rev. C {\bf 82}, 025202 (2010).

\bibitem{mo-tsai}
L.W. Mo and Y.S. Tsai, Rev. Mod. Phys. {\bf 41}, 205 (1969).

\bibitem{geant}
R. Brun {\it et al.}, CERN-DD-78-2-REV, (1978).

\bibitem{exclurad} 
A. Afanasev {\it et al.}, Phys. Rev. D {\bf 66}, 074004 (2002).

\bibitem{peaking}
R. Ent {\it et al.}, Phys. Rev. C {\bf 64}, 054610 (2001).

\bibitem{elastics}
C. Smith and K. Joo, CLAS Analysis Note 2001-018, (2001).

\bibitem{database}
CLAS physics database, \url{http://clasweb.jlab.org/physicsdb}.

\bibitem{adel_saghai} 
R.A. Adelseck and B. Saghai, Phys. Rev. C {\bf 42}, 108 (1990).

\bibitem{deswart} 
J.J. deSwart, Rev. Mod. Phys. {\bf 35}, 916 (1963).

\bibitem{saphir1} 
M.Q. Tran {\it et al.},  Phys. Lett. B {\bf 445}, 20 (1998).

\bibitem{saphir2} 
K.H. Glander {\it et al.}, Eur. Phys. J. A {\bf 19}, 251 (2004).

\bibitem{bradford-cs}
R. Bradford {\it et al.} {\it (CLAS Collaboration)}, Phys. Rev. C {\bf 73}, 035202 (2006).

\bibitem{bennhold}
T. Mart and C. Bennhold, Phys. Rev. C {\bf 61}, 012201 (2000).

\bibitem{saghai}
B. Saghai, AIP Conference Proceedings {\bf 594}, 421 (2001).

\bibitem{sch-sarg}
R.A. Schumacher and M.M. Sargsian, Phys. Rev. C {\bf 83}, 025207 (2011).

\end{thebibliography}
\end{document}